\documentclass[format=acmsmall, review=false, screen=true]{acmart}

\usepackage{graphicx}
\usepackage{tikz} 
\usepackage{booktabs} 
\usepackage{rotating}
\usepackage{array}
\usepackage{makecell}
\usepackage{xcolor}
\usepackage{longtable}
\usepackage{wrapfig}

\usepackage[ruled]{algorithm2e} 

\SetAlFnt{\small}
\SetAlCapFnt{\small}
\SetAlCapNameFnt{\small}
\SetAlCapHSkip{0pt}
\IncMargin{-\parindent}

\setcopyright{acmlicensed}
\acmJournal{CSUR}
\acmYear{2019} \acmVolume{1} \acmNumber{1} \acmArticle{1} \acmMonth{9} \acmPrice{15.00}

\acmDOI{0000001.0000001}

\begin{document}
\title{Scalable Deep Learning on Distributed Infrastructures: Challenges, Techniques and Tools}

\author{Ruben Mayer}
\orcid{0000-0001-9870-7466}
\affiliation{%
  \institution{Technical University of Munich}
  \streetaddress{Boltzmannstr. 3}
  \city{Garching}
  \postcode{85748}
  \country{Germany}}
\email{ruben.mayer@tum.de}
\author{Hans-Arno Jacobsen}
\orcid{000xxx}
\affiliation{%
  \institution{Technical University of Munich}
  \streetaddress{Boltzmannstr. 3}
  \city{Garching}
  \postcode{85748}
  \country{Germany}}
\email{arno.jacobsen@tum.de}

\begin{abstract}
Deep Learning (DL) has had an immense success in the recent past, leading to state-of-the-art results in various domains such as image recognition and natural language processing. One of the reasons for this success is the increasing size of DL models and the proliferation of vast amounts of training data being available. To keep on improving the performance of DL, increasing the scalability of DL systems is necessary. In this survey, we perform a broad and thorough investigation on challenges, techniques and tools for scalable DL on distributed infrastructures. This incorporates infrastructures for DL, methods for parallel DL training, multi-tenant resource scheduling and the management of training and model data. Further, we analyze and compare 11 current open-source DL frameworks and tools and investigate which of the techniques are commonly implemented in practice. Finally, we highlight future research trends in DL systems that deserve further research.
\end{abstract}

%
%


%
%


\maketitle

\vspace{-0.5cm}
\begin{tikzpicture}
\begin{scope}[overlay]
\node[text width=13cm] at ([xshift=-2cm, yshift=-11.5cm]current page.south) {\footnotesize{(c) Owner 2019. This is the author's version of the work. It is posted here for your personal use. Not for redistribution. The definitive Version of Record will be published in ACM Computing Surveys.}};
\end{scope}
\end{tikzpicture}



\section{Introduction}

Deep Learning (DL) has recently gained a lot of attention due to its superior performance in tasks like speech recognition~\cite{6296526, Huang:2014:HPS:2541883.2500887}, optical character recognition~\cite{Borisyuk:2018:RLS:3219819.3219861}, and object detection~\cite{lecun2015deep}. The application of DL poses a tremendous potential in numerous areas like medical image analysis (e.g., breast cancer metastases detection)~\cite{LITJENS201760}, machine translation~\cite{johnson2017google},  image restoration (e.g., automatically colorize grayscale images)~\cite{Iizuka:2016:LCJ:2897824.2925974}, image captioning~\cite{Hossain:2019:CSD:3303862.3295748} (i.e., creating a description of an image), and as agents in reinforcement learning systems that map state-action pairs to expected rewards~\cite{8103164}. In DL, a network of mathematical operators is trained with classified or unclassified data sets until the weights of the model are ready to make correct predictions on previously unseen data. Major companies and open source initiatives have developed powerful DL frameworks such as TensorFlow~\cite{199317} and MXNet~\cite{mxnet} that automatically manage the execution of large DL models developed by domain experts.

One of the driving factors of the success of DL is the scale of training in three dimensions. The first dimension of scale is the size and complexity of the models themselves. Starting from simple, shallow neural networks, with increasing depth and more sophisticated model architectures, new breakthroughs in model accuracy were achieved~\cite{cirecsan2010deep, dean2012large}. The second dimension of scale is the amount of training data. The model accuracy can, to a large extent, be improved by feeding more training data into the model~\cite{4804817, DBLP:journals/corr/abs-1712-00409}. In practice, it is reported that 10s to 100s of Terabyte (TB) of training data are used in the training of a DL model~\cite{186212, 8327042}. The third dimension is the scale of the infrastructure. The availability of programmable highly-parallel hardware, especially graphics processing units (GPUs), is a key-enabler to training large models with a lot of training data in a short time~\cite{cirecsan2010deep, Zhang:2017:PEC:3154690.3154708}. 

Our survey is focused on challenges that arise when managing a large, distributed infrastructure for DL. Hosting a large amount of DL models that are trained with large amounts of training data is challenging. This includes questions of parallelization, resource scheduling and elasticity, data management and portability. This field is now in rapid development, with contributions from diverse research communities such as distributed and networked systems, data management, and machine learning. At the same time, we see a number of open source DL frameworks and orchestration systems emerging~\cite{199317, mxnet_learningsys2016, Peng:2018:OED:3190508.3190517, 222611}. In this survey, we bring together, classify and compare the huge body of work on distributed infrastructures for DL from the different communities that contribute to this area. Furthermore, we provide an overview and comparison of the existing open-source DL frameworks and tools that put distributed DL into practice. Finally, we highlight and discuss open research challenges in this field.

\subsection{Complementary Surveys}

There are a number of surveys on DL that are complementary to ours. Deng~\cite{deng_2014} provides a general survey on DL architectures, algorithms and applications. LeCunn et al. provide a general overview of DL~\cite{lecun2015deep}. Schmidhuber~\cite{SCHMIDHUBER201585} provides a comprehensive survey on the history and technology of DL. Pouyanfar et al.~\cite{Pouyanfar:2018:SDL:3271482.3234150} review current applications of DL.  Luo~\cite{Luo2016} provides a review on hyper-parameter selection strategies in ML training, including training of neural networks.  Those surveys cover general techniques of DL, but are not focused on scalability and distributed systems for DL.

Ben-Nun and Hoefler~\cite{DBLP:journals/corr/abs-1802-09941} provide an analysis of concurrency in parallel and distributed DL training. Chen and Lin~\cite{6817512} provide a survey on DL challenges and perspectives with regard to Big Data (i.e., high data volumes, variety and velocity). Erickson et al.~\cite{Erickson2017} provide a short overview of DL frameworks. Our survey takes a much broader view on distributed DL systems. In particular, we include topics such as resource scheduling, multi-tenancy and data management. Those aspects of scalable DL systems become particularly important when dealing with large models and huge amounts of training data in a shared cluster or cloud environment. Furthermore, we analyze current open-source DL frameworks and tools in depth and relate them to the research on parallel and distributed DL training. This has not been done in the existing surveys. Pouyanfar et al.~\cite{Pouyanfar:2018:SDL:3271482.3234150} analyze and compare DL frameworks, but not with regard to parallelization and distribution.

\subsection{Structure of the Survey}

We structure our survey as follows. In Section~\ref{sec:foundations}, we introduce DL and provide the foundations for the further discussion of DL systems. In Section~\ref{sec:challenges and techniques}, we discuss the challenges and techniques of scalable DL in detail. We cover four important aspects: Distributed infrastructures, parallelization of DL training, resource scheduling and data management. In Section~\ref{sec:comparison}, we analyze and compare 11 open source DL frameworks and tools that put scalable DL into practice. Finally, in Section~\ref{sec:conclusion}, we conclude this survey and provide an outlook on current trends and open problems in the field that deserve further research.

\section{Foundations}
\label{sec:foundations}

\subsection{Context of Deep Learning}
\begin{figure}
\minipage{0.48\textwidth}
    \includegraphics[width=\linewidth]{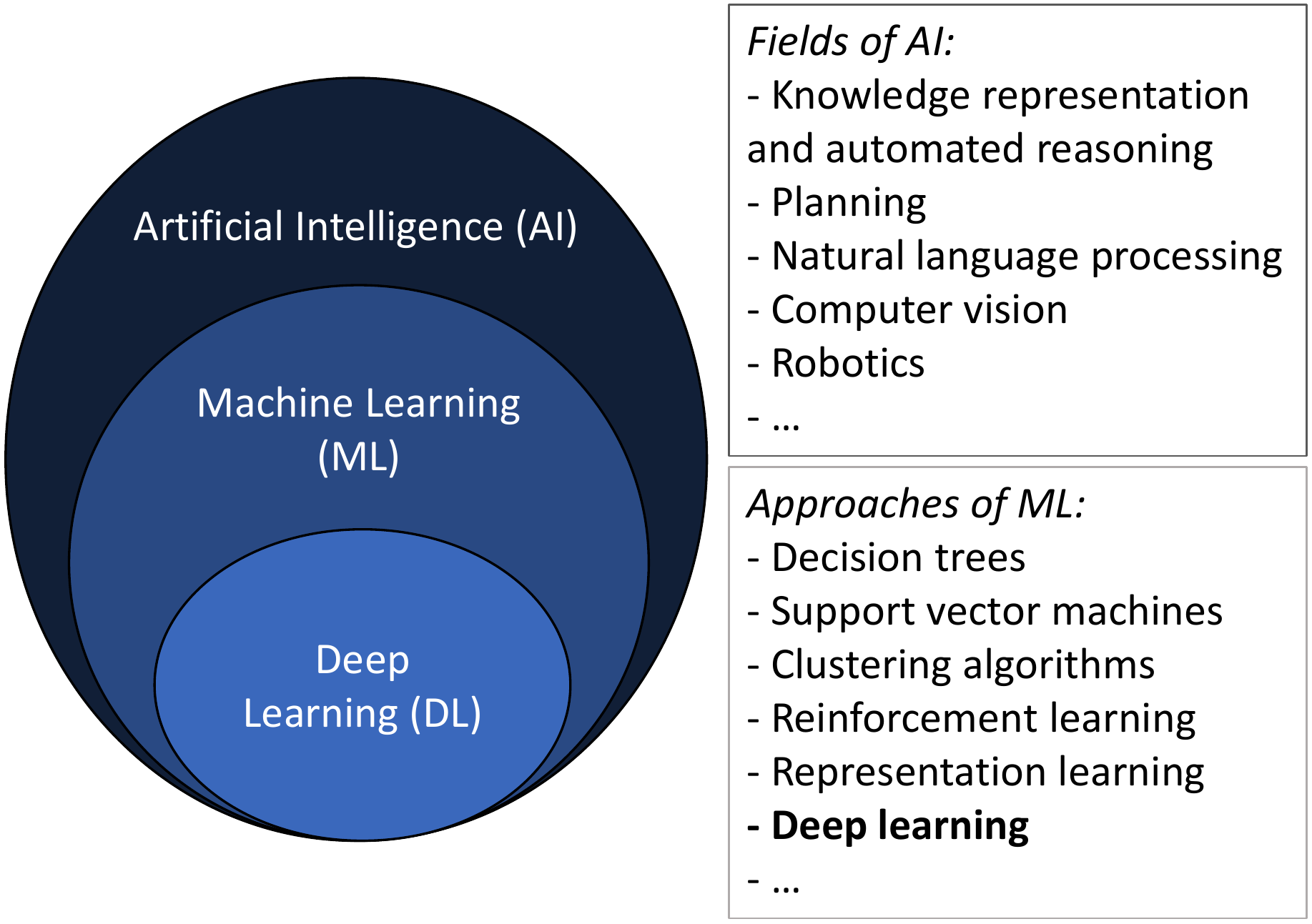}
  \caption{Relationship between AI, ML and DL.}
  \label{fig:aiOverview}
\endminipage\hfill
\minipage{0.46\textwidth}
  \includegraphics[width=\linewidth]{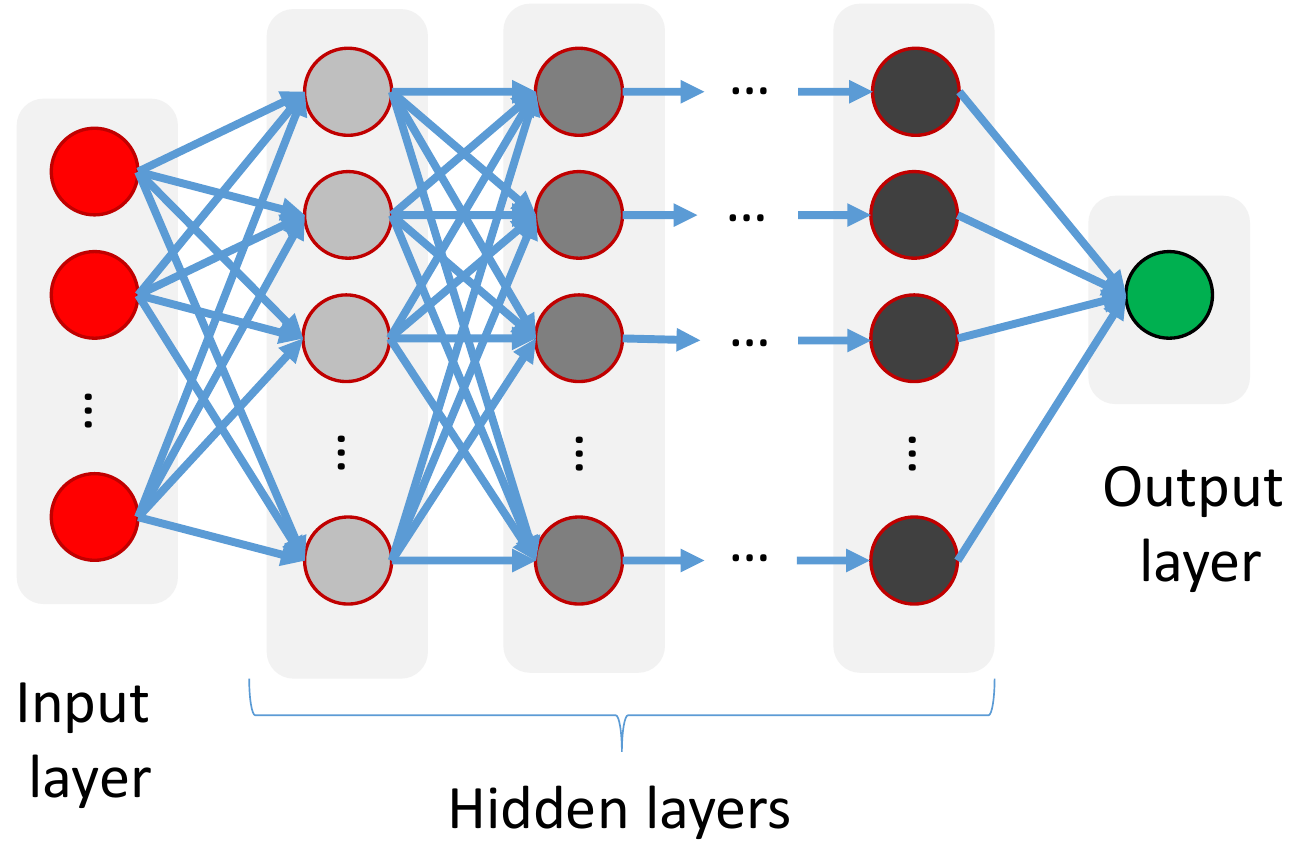}
	\vspace{0.0cm}
  \caption{Schematic of a multi-layer perceptron.}
  \label{fig:mlp}
\endminipage\hfill
\end{figure}

Artificial intelligence (AI) has been a long held vision of building and programming computers in such a way that they can independently (i.e., without human involvement) solve complex problems~\cite{searle1980minds, qai-book}. In the most recent past, immense practical achievements of AI have been made in many different fields, such as knowledge representation and automated reasoning~\cite{NIPS2013_5028}, planning~\cite{NIPS2017_7055}, natural language processing~\cite{8416973}, computer vision~\cite{Szegedy_2016_CVPR}, and robotics~\cite{doi:10.1177/0278364914549607}. Among the methods developed in AI research are cybernetics, symbolic and sub-symbolic, and statistical machine learning (ML). Deep Learning (DL) is a specific approach of ML, which deals with the training of \emph{deep neural networks}. The relationship between AI, ML and DL is visualized in Figure~\ref{fig:aiOverview}.

\subsection{Deep Neural Networks}
\label{sec:Deep Neural Networks}

A neural network (NN) is a network of interconnected \emph{artificial neurons}, which are mathematical functions that transform a set of input signals to an output signal. By layering the neurons and connecting them from an input layer to an output layer, the overall network represents a function $f : x \to y$ that maps the input signals that go into the input layer (layer $1$) to an output signal that leaves the output layer (layer $n$). The goal of $f$ is to approximate a \emph{target function} $f^*$, e.g., a classifier $y = f^*(x)$ that maps an input $x$ to a category $y$. In the \emph{training process}, the set of parameters $\Theta$, i.e., the weights, biases and thresholds, in all of the artificial neurons are adjusted in such a way that the output of $f$ approximates the output of $f^*$ with the best possible accuracy. This is commonly achieved by applying \emph{back-propagation}~\cite{rumelhart1986learning} to the gradient of the loss function w.r.t. the weights of the corresponding layers. There are different gradient descent algorithms applied in DL; a detailed review of gradient descent algorithms is provided by Ruder~\cite{DBLP:journals/corr/Ruder16}.  In the training process, instead of single training samples, \emph{mini-batches} of training data are used in each iteration. This has the advantage of increased parallelism in the training process: The output of the network can be computed for a whole batch of training samples in parallel. However, choosing too large mini-batch sizes may deteriorate the model accuracy and increases the memory footprint of the training process~\cite{DBLP:journals/corr/abs-1804-07612}. The parameters of the training process itself, i.e., the loss function, gradient descent algorithm, activation function, step size (the factor by which the weights are changed toward the gradient), and size of the mini-batches are called \emph{hyper-parameters}.

\subsection{Neural Network Architectures}
The simplest way of organizing a DNN is by using multiple fully-connected layers of neurons, i.e., each neuron in a layer is connected to each neuron in the subsequent layer. This architecture is also referred to as \emph{multi-layer perceptron} (MLP) (cf. Figure~\ref{fig:mlp}). However, MLPs have limitations~\cite{726791, graves2014towards}. First of all, MLPs have a large number of weights, which requires a large number of training samples and occupies a large amount of memory. Second, MLPs are not robust against geometric translations and local distortions of the inputs. For instance, in the detection of hand-written digits from images, the same digit will be written slightly different in different images~\cite{726791}. Third, MLPs are agnostic to the topology of the input, i.e., the order of the input signals is not taken into account. However, in many cases, there is a local structure in the input data. For instance, in images, pixels that are nearby are likely to be correlated~\cite{726791}, and in speech recognition, previous and future context of the input data is particularly relevant to detect a spoken word~\cite{graves2014towards}. To overcome the shortcomings of MLPs, more sophisticated neural network architectures have been proposed. Here, we briefly review the most prominent ones. 

Convolutional neural networks (CNNs)~\cite{726791} introduce convolutional layers and sub-sampling layers. Different from fully-connected layers as in MLPs, convolutional layers are only connected to sub-areas of their respective previous layers, pursuing the concept of \emph{local receptive fields} which is inspired by biology~\cite{hubel1962receptive}. A convolutional layer is composed of multiple planes, where in each plane, all neurons share the same weights (\emph{weight sharing}). Finally, convolutional layers alternate with \emph{sub-sampling} layers to reduce the spacial resolution of the feature map. 
Besides feed-forward networks (where the output of neurons does not loop back to their own input), loop-backs are useful for many use-cases. For instance, in natural language processing, the meaning of one word in a sentence may depend on the meaning of a previously seen word in the same (or even a previous) sentence. To model such phenomena in DL networks, recurrent neural networks (RNNs) have been proposed. Long-short term memory (LSTM) units are special units of an RNN to overcome issues of exploding or vanishing gradients when training RNNs~\cite{doi:10.1162/neco.1997.9.8.1735}. Autoencoders~\cite{Hinton504} are NNs which are used in order to learn efficient encodings (i.e., compressed representations) that extract significant features from the training data. Their architecture consists of an encoder, a code, and a decoder, each consisting of layers of neurons, where the output layer of the network has the same number of neurons as the input layer, but the code, which is exactly between encoding and decoding layers, has much fewer neurons. In generative adversarial networks (GANs) \cite{goodfellow2014generative}, two NNs are aligned with each other, namely, a generative and a discriminative NN. Another recent architecture of NNs are \emph{graph neural networks}~\cite{DBLP:journals/corr/abs-1901-00596}, where graph-structured representations are learned, as opposed to representations in the Euclidian space (as in CNNs).

\section{Distributed Deep Learning}
\label{sec:challenges and techniques}

Training large DL models with vast amounts of training data is a non-trivial task. Often, it is performed in a distributed infrastructure of multiple compute nodes, each of which may be equipped with multiple GPUs. This brings a number of challenges. First of all, the processing resources must be effectively used, i.e., one must avoid stalling of costly GPU resources due to communication bottlenecks. Second, the compute, storage and network resources are typically shared among different users or training processes to reduce costs and provide elasticity (i.e., the cloud computing paradigm~\cite{Armbrust:2010:VCC:1721654.1721672}).  To tackle those challenges in DL, research at the intersection of computing systems and DL is receiving growing attention~\cite{186212, 199317, Cui:2016:GSD:2901318.2901323, Jeong:2018:IED:3190508.3190530, Peng:2018:OED:3190508.3190517, 222611}. This becomes evident with new workshops and conferences arising which particularly focus on DL/ML systems research, such as the \emph{Conference on Systems and Machine Learning} (SysML)\footnote{https://www.sysml.cc}. However, also established communities such as the data management community are turning their attention toward DL/ML systems~\cite{Wang:2016:DMD:3003665.3003669, Kumar:2017:DMM:3035918.3054775}. In this section, we discuss the main directions of DL systems research in depth. We introduce the main research challenges, discuss state-of-the-art approaches, and analyze open research problems that deserve further attention. 

\begin{wrapfigure}{l}{0.54\linewidth}
  \centering
  \includegraphics[width=\linewidth]{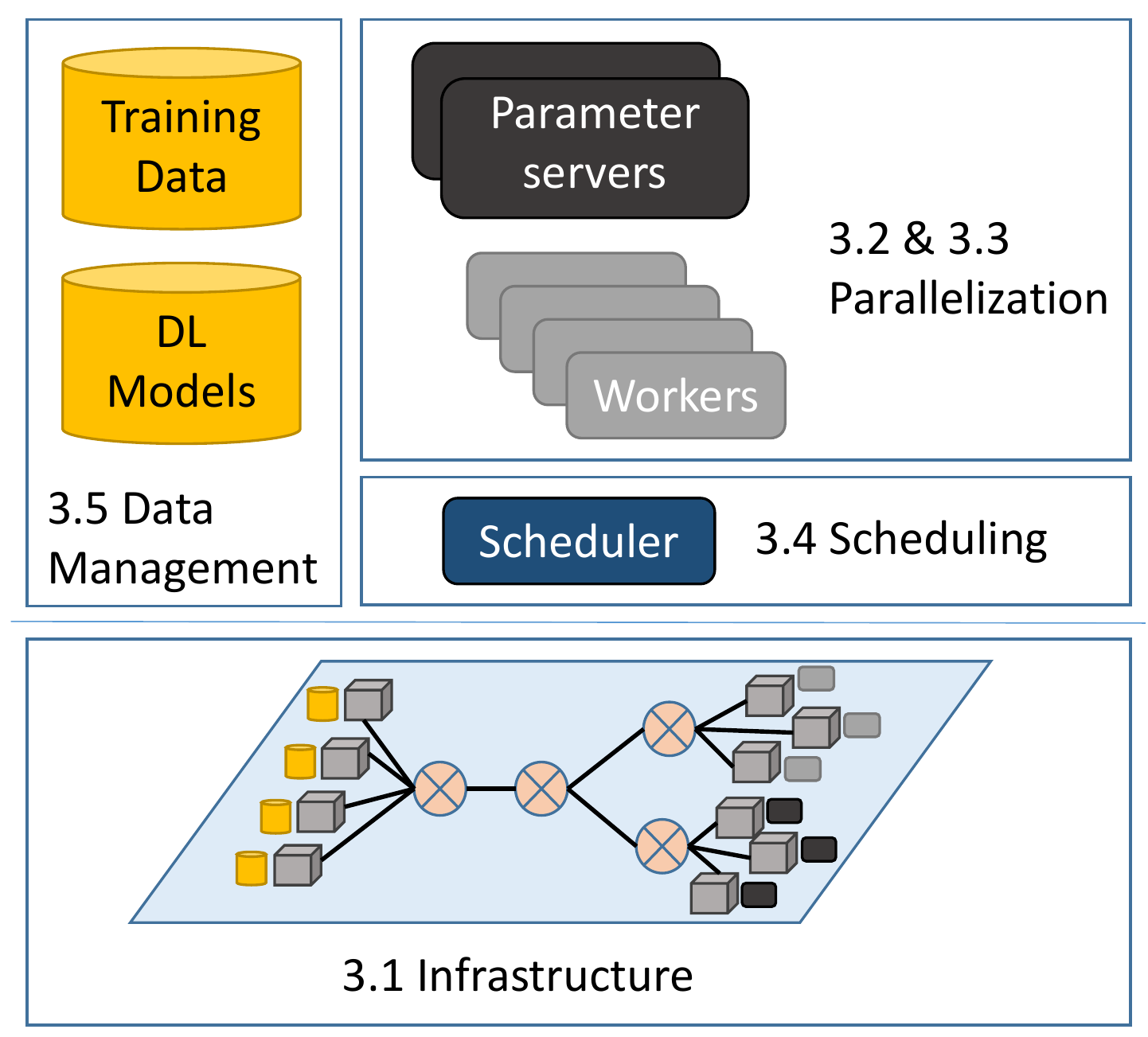}
  \caption{Overview.}
  \label{fig:big_picture}
 \end{wrapfigure}

\textbf{Section Overview.} 
Figure~\ref{fig:big_picture} provides an overview of the topics addressed in this section. On the lowest level, we address the infrastructure used in large DL systems in Section~\ref{sec:Distributed Infrastructures}. We cover recent trends in the hardware being used, networking architectures, as well as low-level software architecture for DL systems. On a higher level, we discuss methods for parallel DL training in Section~\ref{sec:parallelization methods}. In Section~\ref{sec:data-parallel optimizations}, we more specifically discuss challenges and approaches for data-parallel training. To map the components of a parallel DL system to the infrastructure, scheduling is applied. In Section~\ref{sec:scheduling}, we discuss the scheduling problem in single-tenant as well as multi-tenant scenarios. One of the big challenges of large-scale DL is the size of training data and DL models that need to be maintained. In Section~\ref{sec:mod}, we discuss challenges and approaches of data management in DL.

\subsection{Infrastructure}
\label{sec:Distributed Infrastructures}

To understand the challenges on parallelization, scheduling and data management for DL, we first take a deeper look at the infrastructure on which DL training is performed. We divide the existing work into two categories: Hardware innovations and data-center scale infrastructure applied to real DL workloads. While the former can potentially be used on single compute nodes or small clusters, the latter describes how individual hardware components can be composed into a scalable, distributed infrastructure for DL.

\subsubsection{Hardware Components for DL}

While early DL deployments were based on clusters of multi-core CPUs, scalability limitations pushed the efforts to exploiting highly-parallel hardware, and even developing special-purpose hardware dedicated to DL training and serving. 
The performance benefits of GPUs compared to CPU depend on many factors, such as whether the job is processing-bound or memory-bound, the efficiency of the implementation, as well as the hardware itself~\cite{Lee:2010:DGV:1815961.1816021}. Both CPUs and GPUs hardware innovates at a fast pace, which makes comparisons difficult and short-living. Nevertheless, state-of-the-art infrastructures for DL typically comprise GPUs to accelerate the training and inference process. Hardware vendors offer specialized servers and even workstations for DL, such as NVIDIA DGX station~\cite{dgx}.

Besides GPU-centric DL, other forms of hardware acceleration have been proposed, such as field-programmable gate arrays (FPGAs)~\cite{accelerating-deep-convolutional-neural-networks-using-specialized-hardware}.  One strength of FPGAs that is repeatedly mentioned is their capability to make DL training and inference more energy-efficient. NeuFlow by Farabet et al.~\cite{5981829} is one of the first works that tackled the problem of using FPGAs for DL, in particular, for vision systems. Caffeine by Zhang et al.~\cite{7827589} is a hardware and software co-designed library to support CNNs on FPGAs. On the hardware side, it provides a high-level synthesis implementation of an FPGA accelerator for CNNs. In their design, they build upon previously developed methods such as unrolling and pipelining (cf. Zhang et al.~\cite{Zhang:2015:OFA:2684746.2689060}). On the software side, Caffeine provides a driver that allows for easily integrating FPGAs. Caffeine has been integrated into the Caffe DL framework and shows a reduction of energy consumption of up to 43.5x compared to CPU and up to 1.5x compared to GPU execution. Wang et al.~\cite{dlau_wang_et_al} propose a custom FPGA design, called DLAU, to support the training of deep neural networks. One major challenge they had to overcome is the limited memory capacity of FPGAs. They propose tile techniques to partition the training data, along with FIFO buffers and pipelined processing units to minimize memory transfer. In their evaluations, they show that DLAU can train neural networks with up to 10x less energy consumption than GPUs. 
Tensor processing units (TPUs) are application-specific integrated circuits (ASICs) developed by Google that speed-up DL training and inference significantly~\cite{8192463}. TPUs are proprietary and not commercially available, but can be rented via the Google cloud services. 

Besides such more traditional forms of computing architectures that follow the von-Neumann architecture by separating memory and processing units, there are research efforts to develop novel in-memory computing architectures (also called \emph{neuromorphic hardware}~\cite{boybat2018neuromorphic}). Those efforts are inspired by the physiology of the brain, which is very different from the way traditional von-Neumann computing architectures work. Neurostream by Azarkhish et al.~\cite{8038819} is a processor-in-memory solution that is tailored toward training CNNs. However, neuromorphic hardware architectures are still in the experimental stage and not widely available. 

Some papers have highlighted the need for efficient implementations of DL kernels, e.g., by exploiting SIMD (single instruction, multiple data) instructions~\cite{37631, Lee:2010:DGV:1815961.1816021} and awareness of non-uniform memory access (NUMA)~\cite{Roy:2018:NND:3212710.3199605}. This raises the need for re-usable, optimized kernel implementations of the most relevant operations in DNN training. One of the major GPU-specific libraries is cuDNN, a library with DL primitives for GPUs~\cite{DBLP:journals/corr/ChetlurWVCTCS14}. 
The NVIDIA Collective Communications Library (NCCL)~\cite{nccl} provides multi-GPU and multi-node communication primitives and is optimized for PCIe and NVLink high-speed interconnects. DL frameworks often incorporate such low-level libraries to fully exploit the capabilities of the hardware infrastructure.

\subsubsection{Large-scale Infrastructure for DL}

A large-scale DL infrastructure is composed of many inter-connected hardware components that together build a \emph{warehouse-scale computer}~\cite{Barroso:2009:DCI:1643608}. In this subsection, we review current infrastructures as described by organizations that perform very large DL jobs, such as Facebook, Google, and Microsoft, as well as academic research.

Facebook describes its ML infrastructure in a recent paper~\cite{8327042}. They use both CPUs and GPUs for training, and rely on CPUs for inference. To do so, they build specialized CPU-based and GPU-based compute servers to serve their specific needs of training and inference. For training, GPUs are preferred, as they perform better; however, in their data centers, they have abundant capacities of readily-available CPUs, especially during off-peak hours, which they also exploit. For inference, they rely on CPUs, as GPU architectures are optimized for throughput over latency, but latency is a critical factor in inference. Interestingly, for inter-connecting training servers in distributed, data-parallel training, they rely on 50G Ethernet, and forego using specialized interconnects such as RDMA or NCCL~\cite{nccl}. 

Similarly to Facebook, Tencent employs a heterogeneous infrastructure with both CPUs and GPUs. Their deep learning system Mariana~\cite{Zou:2014:MTD:2733004.2733082} consists of three different frameworks that are optimized for different infrastructures and use cases.

Adam is a large-scale distributed system for DL at Microsoft~\cite{186212}. It relies on a large number of commodity hardware CPU-servers to perform DL training. Besides many system-level optimizations, one of the hardware-centric features of Adam is that they partition DL models in such a way that the model layers fit in the L3 cache to improve training performance.

The paper on TensorFlow~\cite{199317}, a scalable ML framework developed by Google, provides some insights into the infrastructure at Google. Overall, Google follows a different approach from Facebook and Microsoft when it comes to the DL infrastructure. First of all, they employ TPUs, which are custom ASICs, as opposed to only using commercial-off-the-shelf (COTS) hardware. Second, they exploit specialized interconnects and use multiple communication protocols, such as gRPC over TCP and RDMA over Converged Ethernet (RoCE)\footnote{RoCE is a network protocol that supports Ethernet as the underlying protocol for remote direct memory access (RDMA).}. Distributed TensorFlow supports communication via the message passing interface (MPI)~\cite{DBLP:journals/corr/VishnuSD16}.

In academic research, exploiting high-performance computing (HPC) infrastructures for DL training is topic with increasing importance. Coates et al.~\cite{Coates:2013:DLC:3042817.3043086} report using a cluster of 16 servers, each equipped with 2 quad-core CPUs and 4 GPUs, being interconnected by Infiniband. Different from Ethernet, Infiniband has high throughput and---more important---extremly low end-to-end latency (in the order of microseconds). Ben-Nun and Hoefler~\cite{DBLP:journals/corr/abs-1802-09941} also observe a trend to move towards HPC infrastructures in DL research.

Summing up, large-scale infrastructures in real-world deployments are highly heterogeneous. They do not only comprise GPU servers, but commonly also CPUs. Overall, we see a certain dominance of COTS hardware, just as it is also the case in other Big Data analytics workloads, such as batch processing~\cite{Dean:2008:MSD:1327452.1327492} and graph processing~\cite{Malewicz:2010:PSL:1807167.1807184}. However, also custom hardware and HPC infrastructure is used, especially at Google and in academic research. In HPC infrastructures, we observe that the DL systems are specialized toward the target infrastructures to increase performance, e.g., regarding the communication protocols like RDMA, NCCL, and MPI.

Performance of distributed infrastructures can be measured, e.g., in terms of throughput, latency and energy consumption. Besides the raw maximum performance of the hardware, another important factor is the communication protocol, e.g., whether RDMA is used. Further important questions are how the hardware components are composed to avoid bottlenecks. Li et al.~\cite{DBLP:journals/corr/abs-1903-04611} have performed a comprehensive performance evaluation of recent GPU interconnects. In terms of energy consumption, Wang et al.~\cite{dlau_wang_et_al} provide evaluations that compare FPGAs to GPUs.

\subsection{Parallelization Methods}
\label{sec:parallelization methods}

DL comes with many possibilities for parallelization. Here, we introduce the three predominant parallelization methods in DL, namely data, model and pipeline parallelism, as well as hybrid forms of parallelism. 

\subsubsection{Data Parallelism}

\begin{wrapfigure}{l}{0.58\linewidth}
  \centering
  \includegraphics[width=\linewidth]{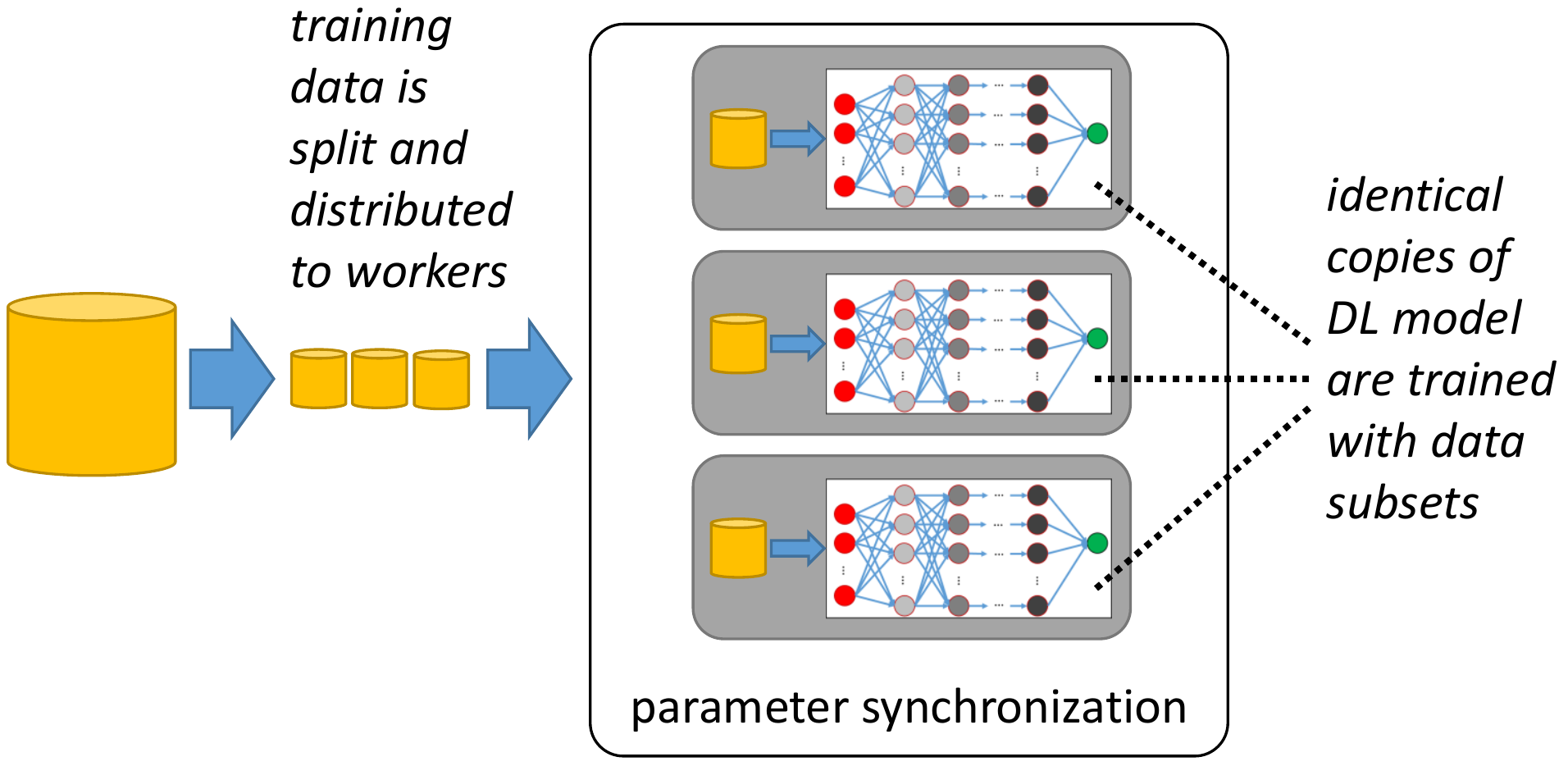}
  \caption{Data parallelism.}
  \label{fig:data_parallelism}
  \vspace{-10pt}
\end{wrapfigure}

In data parallelism, a number of workers (machines or devices, e.g., GPUs) loads an identical copy of the DL model (see  Figure~\ref{fig:data_parallelism}). The training data is split into non-overlapping chunks and fed into the model replicas of the workers for training. Each worker performs the training on its chunk of training data, which leads to updates of the model parameters. Hence, the model parameters between the workers need to be synchronized. There are many challenges in the problem of parameter synchronization. We discuss those challenges and state-of-the-art approaches to tackle them in Section~\ref{sec:data-parallel optimizations}.

The main advantage of data parallelism is that it is applicable to any DL model architecture without further domain knowledge of the model. It scales well for operations that are compute-intensive, but have only few parameters, such as CNNs. However, data parallelism is limited for operations that have many parameters, as the parameter synchronization becomes the bottleneck~\cite{DBLP:journals/corr/abs-1807-05358, DBLP:journals/corr/Krizhevsky14}. This problem could be alleviated by using larger batch sizes; however, this increases data staleness on the workers and leads to poor model convergence. A further limitation of data parallelism is that it does not help when the model size is too large to fit on a single device. It is worth to note that in many data parallel training schemes, it is assumed or required that the training data is independent and identically distributed \emph{(i.i.d.)}, so that parameter updates computed by the parallel workers can simply be summed up in order to compute the new global model parameters~\cite{7239545}.

\subsubsection{Model Parallelism}

\begin{wrapfigure}{r}{0.62\linewidth}
  \centering
  \includegraphics[width=\linewidth]{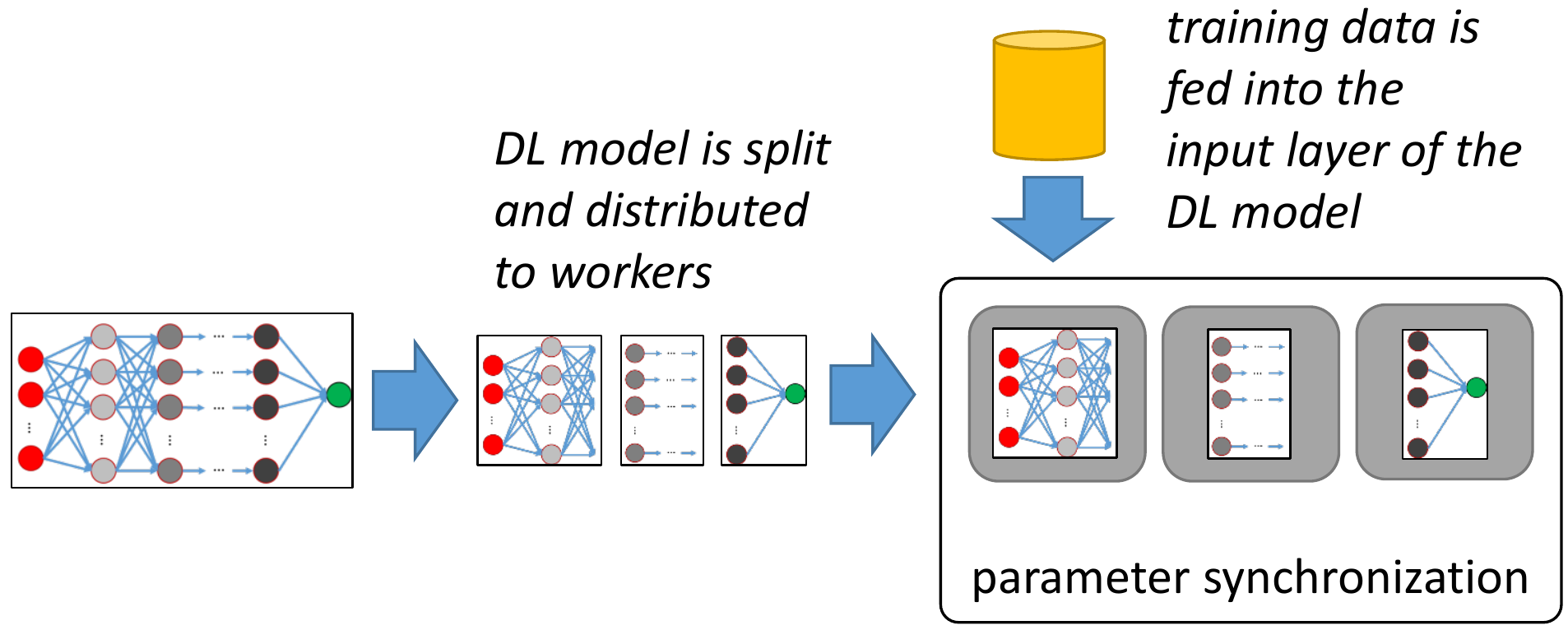}
  \caption{Model parallelism.}
  \label{fig:model_parallelism}
\end{wrapfigure}

In model parallelism, the DL model is split, and each worker loads a different part of the DL model for training (see  Figure~\ref{fig:model_parallelism}). The worker(s) that hold the input layer of the DL model are fed with the training data. In the forward pass, they compute their output signal which is propagated to the workers that hold the next layer of the DL model. In the backpropagation pass, gradients are computed starting at the workers that hold the output layer of the DL model, propagating to the workers that hold the input layers of the DL model.

A major challenge of model parallelism is how to split the model into partitions that are assigned to the parallel workers~\cite{Mayer:2017:TPS:3154842.3154843}. A common approach to find a good model splitting is to use reinforcement learning~\cite{pmlr-v70-mirhoseini17a, mirhoseini2018a}: Starting from some initial partitioning, permutations on that partitioning are performed, and performance is measured (e.g., for one training iteration). In case of an improvement, the permutation is maintained, and further permutations are performed, until the measured performance converges. Streaming rollout~\cite{NIPS2018_7659} is a specialized solution that only works for RNNs.

The main advantage of model parallelism is the reduced memory footprint. As the model is split, less memory is needed for each worker. This is useful when the complete model is too large to fit on a single device. This can be the case when the device consists of specialized hardware such as GPUs or TPUs. The disadvantages of model parallelism are in the heavy communication that is needed between workers. As DL models are hard to be split effectively, there may occur stalling of workers due to communication overhead and synchronization delays. Hence, increasing the degree of model parallelism does not necessarily lead to training speedup~\cite{pmlr-v70-mirhoseini17a}.

\subsubsection{Pipeline Parallelism}

\begin{figure}
  \centering
  \includegraphics[width=0.9\linewidth]{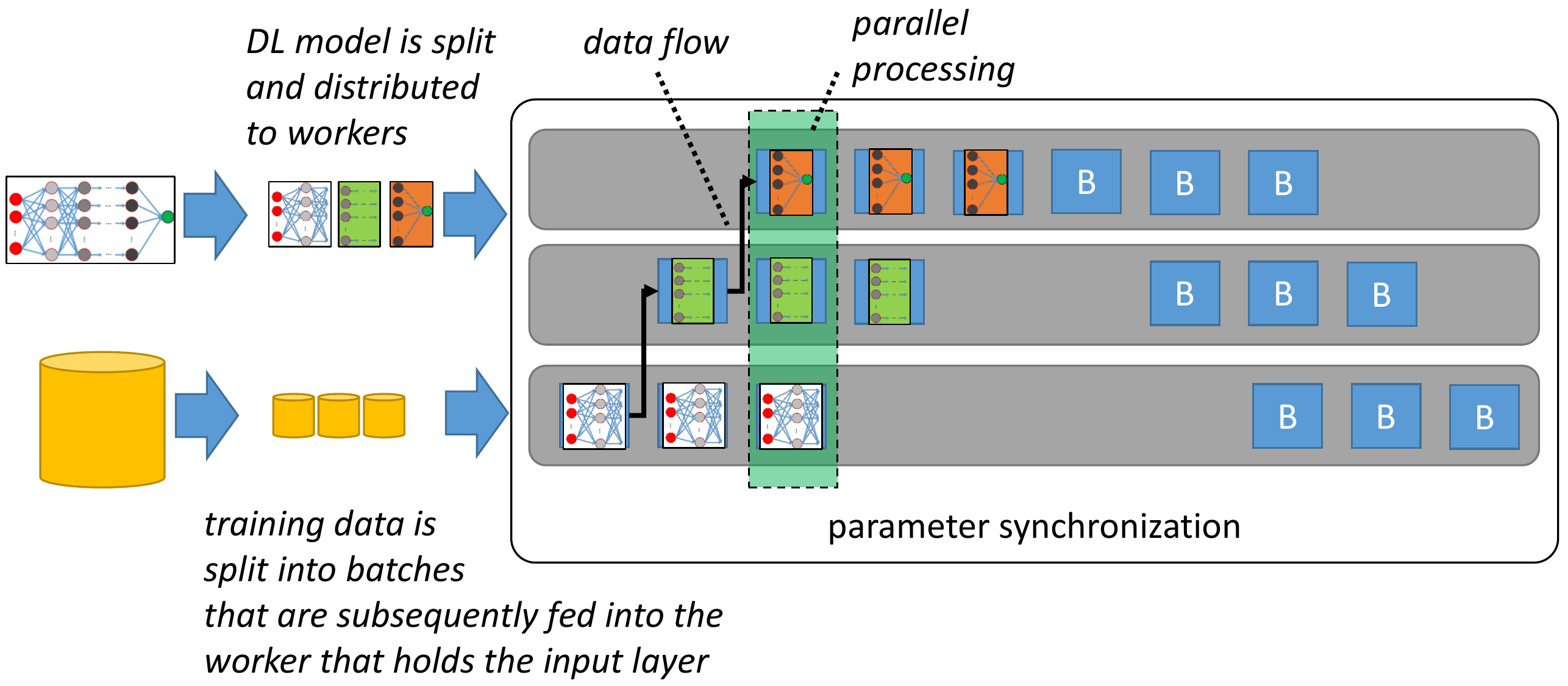}
  \caption{Pipeline parallelism. ``B'' - Backpropagation. Figure adapted from Huang et al.~\cite{DBLP:journals/corr/abs-1811-06965}.} 
  \label{fig:pipeline_parallelism}
 \end{figure}

Pipeline parallelism combines model parallelism with data parallelism. In pipeline parallelism, the model is split and each worker loads a different part of the DL model for training (see Figure~\ref{fig:pipeline_parallelism}). Further, the training data is split into micro batches. Now, every worker computes output signals for a set of micro-batches, immediately propagating them to the subsequent workers. In the same way, in the backpropagation pass, the workers compute gradients for their model partition for multiple micro-batches, immediately propagating them to preceding workers. By streaming multiple micro-batches through the forward and backpropagation pass in parallel, the utilization of workers can be significantly increased compare to pure model parallelism, where only one batch is processed at a time. At the same time, the advantages of model parallelism are maintained, as a single worker does not need to hold the complete model. Current approaches that support pipeline parallelism are GPipe~\cite{DBLP:journals/corr/abs-1811-06965} and PipeDream~\cite{harlappipedream, DBLP:journals/corr/abs-1806-03377}.

\subsubsection{Hybrid Parallelism}

Often, DL models are complex and composed of many different layers that follow a completely different architecture which, in turn, requires different parallelization methods. Hence, hybrid approaches that mix data, model and pipeline parallelism are common.

Mesh-TensorFlow~\cite{shazeer2018mesh} is a language extension of TensorFlow that allows for combining data parallelism and model parallelism.  In Mesh-TensorFlow, tensors can be split across a ``mesh'' of processors (such as CPUs, GPUs or TPUs). To achieve data parallelism, data is split into shards; to achieve model parallelism, tensors are split along any of their attributes.

There are a couple of papers that propose optimizations of parallelization that are manually designed by domain experts. Krizhevsky~\cite{DBLP:journals/corr/Krizhevsky14} proposed to apply data parallelism for convolutional and pooling layers, as those layers are compute-heavy and only have few parameters, and model parallelism for fully-connected layers, as they are light in computation, but have many parameters. In Google's Neural Machine Translation System (GNMT)~\cite{DBLP:journals/corr/WuSCLNMKCGMKSJL16} that powers Google Translate, they apply data parallelism, but combine it with hand-crafted model parallelism for each model replica.

Beyond manually designed hybrid models, recently, automated optimization approaches have been developed. Jia et al.~\cite{pmlr-v80-jia18a} propose ``layer-wise'' parallelization. For each layer of a DNN, an optimal parallelization method is chosen along the tensors' dimensions at the layer. To do so, they employ a cost model and a graph search algorithm on a reduced graph that models the solution space. FlexFlow by Jia et al.~\cite{DBLP:journals/corr/abs-1807-05358} is an automatic parallelization optimizer that employs an execution simulator. It optimizes parallelism across four dimensions, referred to as the \emph{SOAP space}: the \underline{s}ample, \underline{o}peration, \underline{a}ttribute and \underline{p}arameter dimension. The sample dimension refers to batches of training data and corresponds to data parallelism. The operation dimension refers to artificial neurons, the attribute dimension refers to the attributes of the tensors, and the parameter dimension refers to the weights and other model parameters. Together, the operation, attribute and parameter dimensions correspond to model parallelism~\cite{pmlr-v80-jia18a}.

\subsection{Optimizations for Data Parallelism}
\label{sec:data-parallel optimizations}
Parameter synchronization in data-parallel DL systems poses three major challenges. The first challenge is \emph{how} to synchronize the parameters. Should the workers synchronize via a centralized architecture or in a decentralized manner? The second challenge is \emph{when} to synchronize the parameters. Should the workers be forced to synchronize after each batch, or do we allow them more freedom to work with potentially stale parameters? The third challenge is how to \emph{minimize communication overhead} for synchronization. 

\subsubsection{System Architecture}
\label{sec:data-parallel architecture}

\begin{figure}
  \begin{minipage}{.41\textwidth}
  \centering
  \includegraphics[width=\linewidth]{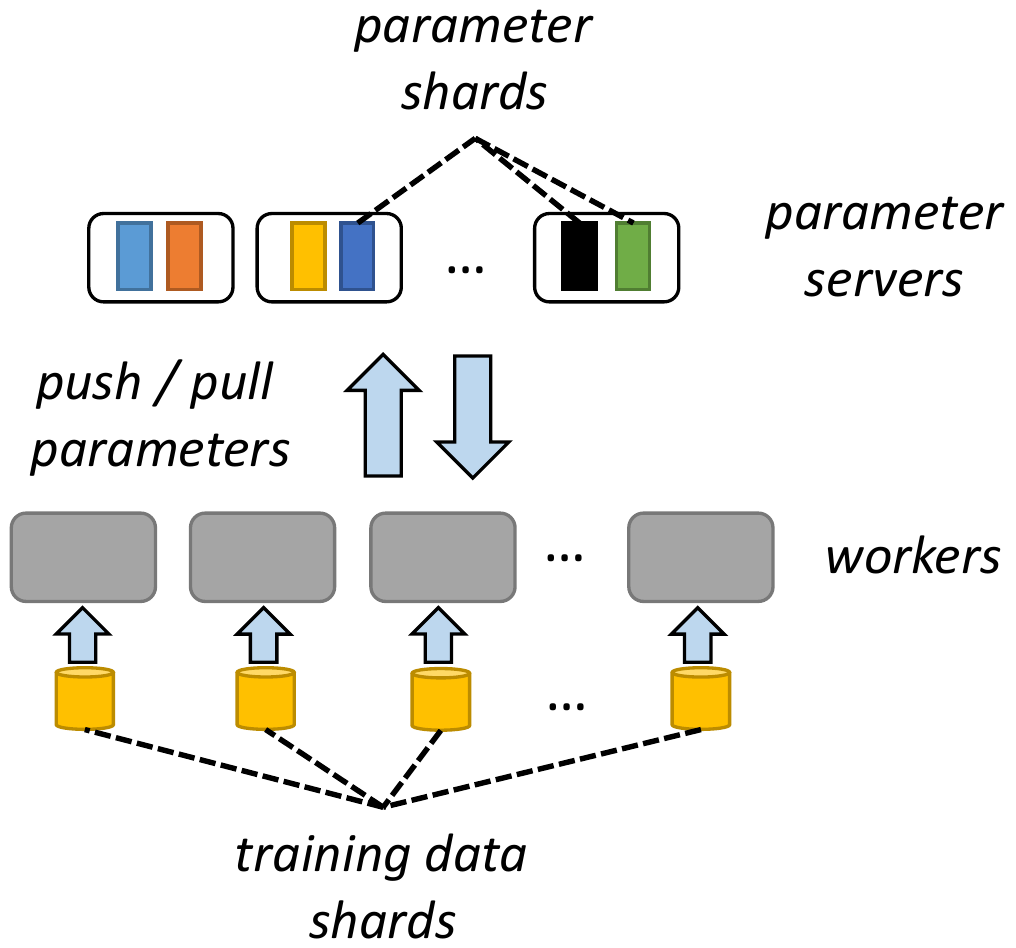}
  \caption{Parameter server architecture.}
  \label{fig:ps_architecture}
  \end{minipage}\quad 
    \begin{minipage}{.43\textwidth}
  \centering
 \includegraphics[width=\linewidth]{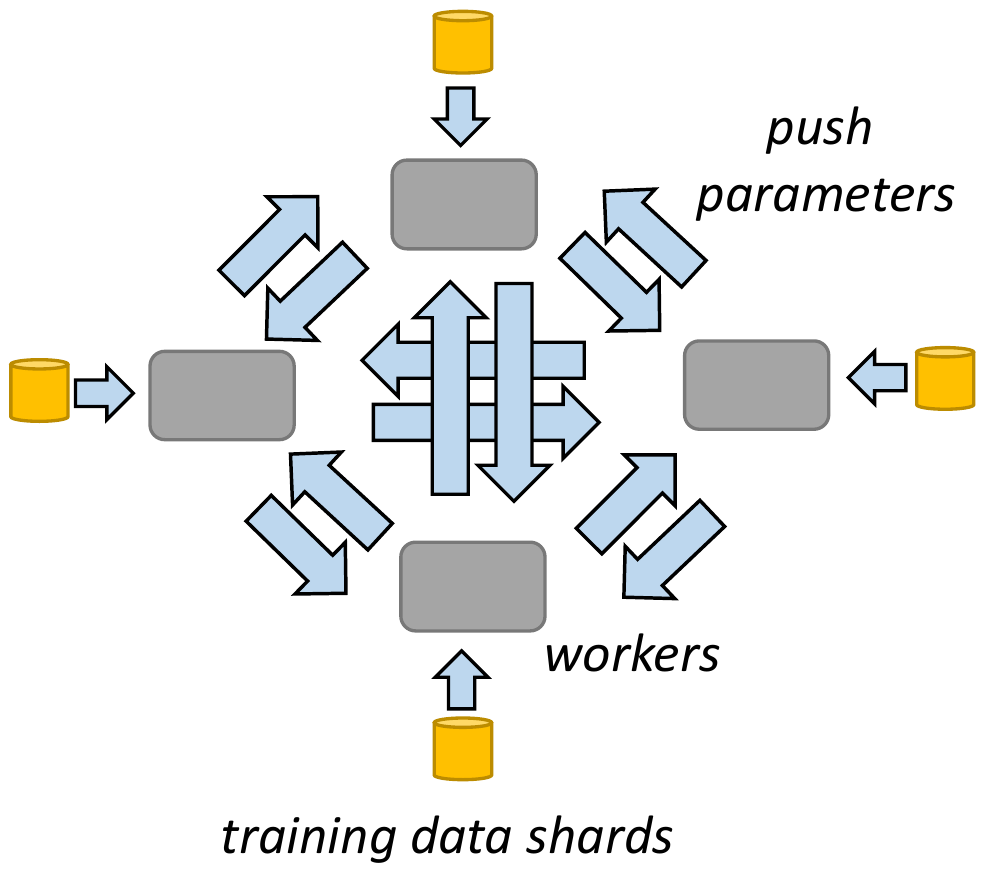}
  \caption{All-reduce architecture.}
\label{fig:dec_architecture}
  \end{minipage}
\end{figure}

The system architecture describes \emph{how} the parameters of the different workers are synchronized. One of the major challenges is to provide a scalable system architecture than can deal with a large number of parallel workers that regularly update the DL model as well as receive an updated view of the model for further training. The second challenge is to keep the system easy to configure, i.e., it should be possible to yield good performance without needing extensive parameter tuning. The third challenge is to exploit lower-level primitives, e.g., communication primitives such as offered by NCCL, in an optimal way.

\textit{(1) Centralized.} In the (logically) centralized architecture, workers periodically report their computed parameters or parameter updates to a (set of) \emph{parameter server(s) (PSs)} (see Figure~\ref{fig:ps_architecture}). Roots of the PS architecture go back to the blackboard architecture~\cite{Smola:2010:APT:1920841.1920931} and MapReduce~\cite{Dean:2008:MSD:1327452.1327492}, as Alex Smola reports~\cite{smola_on_ps}. The PS architecture is the most prominent architecture of data parallel DL systems. A common approach is to use sharding of the model parameters and distribute the shards on multiple PSs which then can be updated in parallel~\cite{dean2012large}. 
Among the systems that use a parameter server architecture are GeePS~\cite{Cui:2016:GSD:2901318.2901323}, DistBelief~\cite{dean2012large}, TensorFlow~\cite{199317}, Project Adam~\cite{186212}, Poseidon~\cite{Zhang:2017:PEC:3154690.3154708}, SINGA~\cite{Ooi:2015:SDD:2733373.2807410}, SparkNet~\cite{moritz2015sparknet} and the system by Yan et al.~\cite{Yan:2015:PMS:2783258.2783270}.

\textit{(2) Decentralized.} The decentralized architecture works without a PS. Instead, the workers exchange parameter updates directly via an \emph{allreduce} operation (see Figure~\ref{fig:dec_architecture}). In doing so, the \emph{topology} of the workers plays an important role. A fully connected network, where each worker communicates with each other worker, has a communication cost that is in $\mathcal{O}(n^2)$ with $n$ workers, so that communication becomes a bottleneck. A common alternative is to employ a ring topology (referred to as \emph{ring-allreduce}). Horovod~\cite{DBLP:journals/corr/abs-1802-05799} from Uber uses NCCL to implement ring-allreduce. Baidu had one of the first proposals of using ring-allreduce for data parallel DL training~\cite{baidu-allreduce}. The multi-GPU framework in Tencent's Mariana DL system~\cite{Zou:2014:MTD:2733004.2733082} employs a similar linear topology for parameter exchange across workers. Other topologies that have been proposed are ``Butterfly''~\cite{doi:10.1137/1.9781611972832.87}, a tree~\cite{JMLR:v15:agarwal14a}, and a graph that is built based on a Halton sequence~\cite{Li:2015:MDD:2741948.2741965}. Wang et al.~\cite{Wang:2014:STC:2637166.2637231} propose a parameter sharing protocol that allows for arbitrary loop-free worker topologies that can also be dynamically changed at system run-time. The main drawback of alternative topologies, different from the fully connected topology, is that the propagation of parameter updates to all workers needs more time, as there may be multiple hops between a pair of workers.

The topology of the workers is not the only knob to reduce network load. Ako by Watcharapichat et al.~\cite{Watcharapichat:2016:ADD:2987550.2987586} employs a fully connected network of workers, but partitions the gradients that are exchanged between workers (\emph{partial gradient exchange}). In each round of synchronization, each worker only sends a single partition of the gradients to every other worker; in particular, it may send different partitions to different workers. Clearly, the communication overhead depends both on the size of a partition (which itself depends on the number of partitions) as well as on the number of workers. The number of partitions is adapted automatically in such a way that the network bandwidth remains constant independently of the number of workers. 

\emph{Comparison to centralized architecture.} \quad The advantages of the decentralized architecture compared to the centralized one are the following, according to Li et al.~\cite{Li:2015:MDD:2741948.2741965}. By using the decentralized architecture, one avoids the need to deal with the inconveniences of implementing and tuning a parameter server. This is not only a matter of the complexity of the system code, but also eases the deployment. One does not need to plan which resources to allocate for the parameter servers and for the workers. A further advantage is that fault tolerance can be achieved more easily, because there is no single point of failure such as the parameter server. When a node in the decentralized architecture fails, other nodes can easily take over its workload and the training proceeds without interruptions. Heavy-weight checkpointing of the parameter server state is not necessary.

The decentralized architecture also has disadvantages. First and foremost, communication in the decentralized architecture increases quadratically with the number of workers, if no counter-measures are taken. As discussed above, those counter-measures, such as changing the topology or partitioning the gradients, induce new complexities and trade-offs. Overall, there is no silver bullet for the problem of synchronizing parallel parameter updates.

A case study by Lian et al.~\cite{lian2017can} indicates that the decentralized architecture can, under certain conditions, perform better than the centralized architecture if the communication network is slow. However, their study is limited to synchronous parameter updates and the centralized architecture they compare to employs only a single parameter server. In such a setting, the network connecting the single central parameter server quickly becomes the bottleneck. Similar results have been reported by Iandola et al.~\cite{Iandola_2016_CVPR} who also prefer a tree-structured allreduce architecture to a single parameter server.

Both centralized and decentralized learning are widely implemented in open source DL frameworks. Some frameworks, such as TensorFlow and MXNet, even support both. In TensorFlow, the decentralized architecture is applied to training on a single compute node with multiple GPUs, as efficient allreduce implementations such as NCCL allreduce can be used. On the other hand, the centralized architecture is applied to multi-node training~\cite{tf-distribute-doc}.

\textit{(3) Federated.} Both the centralized and the decentralized architecture assume a controlled environment (such as a data center), a balanced and i.i.d. distribution of the training data to the workers, and a network with homogeneous, high bandwidth. In contrast to this, federated learning~\cite{45648} evolves around a scenario where the training data is kept locally on users' mobile devices, and a global model is trained based on updates that the users compute on their local devices. That way, training data, which may contain privacy-sensitive information, can be completely kept locally, which can also decrease the bandwidth requirements between the mobile devices and the central data center. 

The low and asynchronous bandwidth (i.e., the uplink is usually much slower than the downlink) of a mobile device's Internet connection makes it impossible to repeatedly upload the updated parameters of a large model to a centralized parameter server or to decentralized peer nodes. Kone\u{c}n\'{y} et al.~\cite{45648} study different forms of parameter sampling and compression to mitigate this problem. McMahan et al.~\cite{44822} propose the \emph{federated averaging} algorithm for reducing the parameter updates. Their algorithm is round-based: In each round, a fraction of the clients is selected. Each selected client computes the gradient of the loss function over all the training data that it holds. To reach convergence, it is important that the model instances on the client start from the same random initialization. Finally, a central server aggregates the gradients from the selected clients. In a comparative performance study by Nilsson et al.~\cite{Nilsson:2018:PEF:3286490.3286559}, the authors show that federated averaging is the best algorithm for federated learning, and is practically equivalent to the centralized architecture when i.i.d. training data is used. However, in the non-i.i.d. case, the centralized approach performs better than federated averaging.

Federated learning is still in an early stage and is not widely supported yet in open source DL frameworks. Recently, first tools for federated learning were made available. TensorFlow Federated~\cite{tf-federated} is a simulator for experimenting with federated ML. PySyft~\cite{DBLP:journals/corr/abs-1811-04017, pysyft-github} is a Python library that enables privacy-preserving federated learning within PyTorch. In particular, PySyft applies \emph{differential privacy} methods~\cite{Abadi:2016:DLD:2976749.2978318} to federated learning to prevent that sensitive information about the training data can be extracted from the model.

\newcolumntype{v}[1]{%
	>{\begin{turn}{90}\begin{minipage}{#1}\raggedright\hspace{0pt}}l%
	<{\end{minipage}\end{turn}}%
	}


\begin{table}[]
	\footnotesize
	\begin{tabular}{|l|p{1.3cm}||c|c|c||c|c|c||c||p{2.6cm}|c|c|}
		\hline  
		&  & \multicolumn{3}{c||}{\thead{Synchronization \\ Model}}  & \multicolumn{3}{c||}{\thead{System \\ Architecture}} &   &  \\
	\thead{Ref.} & \thead{Name} & \thead{Sync.}  & \thead{Bound. \\ Async.}  & \thead{Async.} & \thead{Centra- \\ lized}  & \thead{Decen- \\ tralized}  & \thead{Fede- \\ rated}  & \thead{Year}   & \thead{Main Concepts} \\
		\hline 
		\cite{recht2011hogwild} & Hogwild  &  &  & x &  & x & & 2011  & Lock-free updates \\  \hline
		\cite{dean2012large} & Downpour SGD  &  &  & x & x & & & 2012  & Parameter sharding, asynchronous SGD \\  \hline
		\cite{181983} & Cipar et al.  &  & x &  & x & & & 2013  & Introduces Stale Synchronous Parallel (SSP) \\  \hline
		\cite{noel2014dogwild} & Dogwild  &  &  & x &  & x & & 2014  & Distributed Hogwild~\cite{recht2011hogwild} \\  \hline
		\cite{Cui:2014:EBS:2643634.2643639} & Cui et al.  &  & x &  & x & & & 2014  & Applies SSP~\cite{181983} \\  \hline
		\cite{186214} & Li et al.  & x & x & x & x & & & 2014  & Flexible consistency \\ \hline
		\cite{Dai:2015:HDM:2887007.2887019} & Dai et al.  &  & x &  & x & & & 2015  & Introduces Eager SSP \\ \hline
		\cite{Li:2015:MDD:2741948.2741965} & MALT  &  &  & x &  & x & & 2015  & Shared memory abstraction \\  \hline
		\cite{7837887} & Hogwild++  &  &  & x &  & x & & 2016  & NUMA-aware Hogwild~\cite{recht2011hogwild} \\  \hline
		\cite{Cui:2016:GSD:2901318.2901323} & GeePS  &  x & x  & x & x &  & & 2016  & GPU-specialized PS \\  \hline
		\cite{Jiang:2017:HDP:3035918.3035933} & Jiang et al.  &  & x &  & x &  & & 2017  & Dynamic learning rates on SSP~\cite{181983} \\  \hline
		\cite{Wang:2018:ASP:3274808.3274828} & A-BSP  & x &  &  & x & x & & 2018  & Aggressive synchronization \\  \hline
		\cite{DBLP:journals/corr/abs-1901-02244} & CROSS-BOW  & x &  &  &  & x & & 2019  & Synchronous model averaging \\  \hline
		\cite{47976} & Bonawitz et al.  & x &  &  & x &  & x & 2019  & Synchronous federated learning\\  

\hline 
		
	\end{tabular} 
	\caption{Categorization of approaches on parameter synchronization in data-parallel training. }
	\label{tab:synchronizationApproaches}
	\vspace{-15pt}
\end{table}

\subsubsection{Synchronization}
\label{sec:data-parallel parameter synchronization}

The question \emph{when} to synchronize the parameters between the parallel workers has received a lot of attention.  The main challenge in parameter synchronization is to handle the trade-off between the potential loss in training quality or convergence speed when workers perform training on a stale DL model and the synchronization cost to update the DL model on the workers. Overall, there are three different main approaches: Synchronous, bounded asynchronous, and asynchronous training. Table~\ref{tab:synchronizationApproaches} provides an overview and categorization of the most relevant publications.

\emph{(1) Synchronous.}\  In synchronous training, after each iteration (processing of a batch), the workers synchronize their parameter updates. Such a strict model can be implemented by well-known abstractions such as the Bulk Synchronous Parallel (BSP) model~\cite{Valiant:1990:BMP:79173.79181}, which are in many cases already available in data analytics platforms such as Hadoop / MapReduce~\cite{Dean:2008:MSD:1327452.1327492}, Spark~\cite{180560, JMLR:v17:15-237} or Pregel~\cite{Malewicz:2010:PSL:1807167.1807184}. The advantage of strict synchronization is that reasoning about the model convergence is easier. However, strict synchronization makes the training process prone to the \emph{straggler problem}, where the slowest worker slows down all others~\cite{181983}.

GeePS~\cite{Cui:2016:GSD:2901318.2901323} by Cui et al. is a parameter server implementation that is tailored to GPUs. This includes a couple of optimizations such as pre-built indexes, caching, data staging and memory management. While GeePS supports synchronous, bounded asynchronous and asynchronous parameter synchronization, it is designed to minimize the straggler problem on GPUs, and hence, achieves best convergence speed when using the synchronous approach. Wang et al.~\cite{Wang:2018:ASP:3274808.3274828} propose an \emph{aggressive synchronization} scheme that is based on BSP, named A-BSP. Different from BSP, A-BSP allows the fastest task to fetch current updates generated by the other (straggler) tasks that have only partially processed their input data. The authors have implemented A-BSP both on  Spark~\cite{180560, JMLR:v17:15-237} as well as on the Petuum system~\cite{7239545}. CROSSBOW~\cite{DBLP:journals/corr/abs-1901-02244} by Koliousis et al. introduces \emph{synchronous model averaging} (SMA). In SMA, data-parallel workers access a global average model in order to coordinate with each other. In particular, the workers independently train their model replica on their respective shard of the training data, but \emph{correct} their model parameters according to the difference of their local models to the global average model. Bonawitz et al.~\cite{47976} discuss a system design that is tailored to synchronous training for federated learning. The main challenges they address are how to deal with fluctuating device availability and churn, interrupted connectivity and limited device capabilities. To solve these challenges, they propose to employ a centralized architecture with a parameter server. The training process is divided into subsequent rounds; after each round, locally computed gradient updates are collected from the participating devices and aggregated on the parameter server using federated averaging. By selecting a new set of devices for participation in each training round, the parameter server can balance the load among devices and can flexibly react on dynamics such as device churn.

Synchronous training is implemented in a wide range of open-source DL frameworks, such as TensorFlow~\cite{199317, tf-distribute-doc} and MXNet~\cite{mxnet_learningsys2016, mxnet-distribute-doc}. It is especially suitable for parallel training on a single, multi-GPU compute node, where communication delays are small and computational load is balanced, such that the straggler problem is not significant~\cite{mxnet-github-discussion, tf-distribute-doc}.

\emph{(2) Bounded asynchronous.}\  Asynchronous training makes use of the approximate nature of DL training. Recall, that DL models are mathematical functions that approximate the target function $f^*$ as good as possible (cf. Section~\ref{sec:Deep Neural Networks}). Hence, small deviations and non-determinism in the training process do not necessarily harm the model accuracy. This is different from ``strict'' problems in data analytics, such as database queries, which are required to return a deterministic result. In bounded asynchronous training, workers may train on stale parameters, but the staleness is bounded~\cite{181983}. Bounded staleness allows for a mathematical analysis and proof of the model convergence properties. The bound allows the workers for more freedom in making training progress independently from each other, which mitigates the straggler problem to some extent and increases throughput.

Cipar et al. introduced the Stale Synchronous Parallel (SSP) model~\cite{181983}. Different from the BSP model, SSP allows for bounded staleness of the workers, i.e., there may be a delay between a worker updating the parameters and the effects of that update being visible to other workers. This delay is given in terms of a number of iterations. A follow-up paper by Cui et al.~\cite{Cui:2014:EBS:2643634.2643639} proposes an implementation of SSP for ML jobs. Dai et al.~\cite{Dai:2015:HDM:2887007.2887019} perform a theoretical analysis of SSP, comparing it against a theoretically optimal (but practically not implementable) approach. In the course of their analysis, they propose Eager SSP (ESSP), which is a novel implementation of the SSP model. In ESSP, workers eagerly pull updates from the parameter servers, as opposed to SSP where updates are only pulled when the worker state becomes too stale. ESSP is implemented in the Petuum system~\cite{7239545}. The parameter server by Li et al.~\cite{186214} has a flexible consistency model that also supports bounded delays. Jiang et al.~\cite{Jiang:2017:HDP:3035918.3035933} propose to use dynamic learning rates on top of SSP to account for heterogeneous workers. Depending on a worker's speed, its learning rate is adapted such that stale updates have a less significant effect on the global parameters than fresh updates. 

The bounded asynchronous model is not widely implemented in DL frameworks, as Zhang et al.~\cite{8416283} notice. Li~\cite{mxnet-github-discussion} noted in a Github discussion that SSP was not implemented in MXNet, because the observed delays were only small due to the uniform performance of GPU-intensive operations, such that the benefits of SSP were not significant enough. There are some exceptions. The Parallel ML System (PMLS) uses Bösen~\cite{Wei:2015:MCC:2806777.2806778}, a bounded-asynchronous parameter server. However, PMLS and Bösen are no longer actively developed. CNTK~\cite{Seide:2016:CMO:2939672.2945397} implements blockwise model update and filtering (BMUF)~\cite{7472805}, a variant of bounded asynchronous training. Petuum, which is a commercial product, implements the bounded asynchronous model~\cite{7239545}.

\emph{(3) Asynchronous.}\  In asynchronous training, workers update their model completely independently from each other. There are no guarantees on a staleness bound, i.e., a worker may train on an arbitrarily stale model. This makes it hard to mathematically reason about the model convergence. However, on the other hand side, it provides the workers the greatest possible flexibility in their training process, completely avoiding all straggler problems. 

Hogwild~\cite{recht2011hogwild} by Recht et al. is an asynchronous implementation of parallel SGD. The parameter update scheme of Hogwild grants the workers access to shared memory without any locks, i.e., workers can overwrite each other's updates of the model parameters. This seems dangerous due to the lost update problem: New model parameters written by one worker could directly be overwritten by another worker and, hence, would not have any effect. However, the authors show that as long as the updates of the single workers only modify small parts of the model, Hogwild achieves nearly optimal convergence. By foregoing locks, Hogwild performs by an order of magnitude faster than update schemes that lock the model parameters before each update. The Hogwild scheme has been successfully applied to the training of neural networks~\cite{ParallelizationofNeuralNetworkTrainingforNLPwithHogwild}. Dogwild~\cite{noel2014dogwild} by Noel and Osindero is a distributed implementation of Hogwild. The authors report that using UDP congested the network stack, while using TCP did not fully utilize the communication bandwidth and also caused latency spikes, so that they use raw sockets instead. Hogwild++~\cite{7837887} by Zhang et al. is an adaptation of Hogwild to NUMA-based memory architectures. 
Downpour SGD~\cite{dean2012large} by Dean et al. is an asynchronous SGD procedure tailored to large clusters of commodity machines. Among the main concepts of Downpour SGD are the sharded parameter server and the application of adaptive learning rates~\cite{Duchi:2011:ASM:1953048.2021068}. Different from Hogwild, which is lock-free, Downpour SGD uses lock-guarded parameter increments.
MALT~\cite{Li:2015:MDD:2741948.2741965} by Li et al. is an asynchronous ML framework that follows the decentralized architecture. It provides a shared memory abstraction for the workers that provides a scatter/gather interface as well as a higher-level vector object library.

The same as synchronous training, asynchronous training is well-established; there are many implementations in current open source DL frameworks, such as TensorFlow~\cite{tf-distribute-doc}, MXNet~\cite{mxnet-distribute-doc}, CNTK~\cite{cntk-distribute-doc} and PyTorch~\cite{pytorch-distribute-doc}.

\newcolumntype{v}[1]{%
	>{\begin{turn}{90}\begin{minipage}{#1}\raggedright\hspace{0pt}}l%
	<{\end{minipage}\end{turn}}%
	}


\begin{table}[]
	\footnotesize
	\begin{tabular}{|l|p{1.3cm}||c|c|c||c|c|c||c|c|c||c|c|}
		\hline 
		&  & \multicolumn{3}{c||}{\thead{Communication \\ Optimization}}  & \multicolumn{3}{c||}{\thead{Synchronization \\ Model}}  & \multicolumn{3}{c||}{\thead{System \\ Architecture}} &   \\
	\thead{Ref.} & \thead{Name} & \thead{Preci- \\ sion}  & \thead{Com-\\press.} & \thead{Comm. \\ Sched.}  &  \thead{Sync.}  & \thead{Bound. \\ Async.}  & \thead{Async.}   & \thead{Centra- \\ lized}  & \thead{Decen- \\ tralized}  & \thead{Fede- \\ rated}    & \thead{Year}  \\
		\hline 
		\cite{seide-compress-gradients} & Seide et al.  &  & x &  & x &  & &   & x &  & 2014 \\  \hline
		\cite{186214} & Li et al.  &  & x &  & x&x &x & x  &  & & 2014\\ \hline
		\cite{Gupta:2015:DLL:3045118.3045303} & Gupta et al.  & x &  &  & x & x & x &  x & x & x & 2015 \\  \hline
		\cite{Wei:2015:MCC:2806777.2806778} & B\"{o}sen  &  &  & x  &  & x & & x & &   & 2015 \\  \hline
		\cite{190634} & MLNet  &  &  & x  & x & x & &  x & &   & 2015 \\  \hline
		\cite{DBLP:journals/corr/ZhouNZWWZ16} & DoReFa-Net  & x &  &  & x & x & x & x  & x & x & 2016 \\  \hline
		\cite{qsgd-alistarh} & QSGD  &  & x &  & x & x & x &   & x &  & 2017 \\  \hline
		\cite{wen2017terngrad} & TernGrad  &  & x &  & x &  & & x  &  &  & 2017 \\  \hline
		\cite{lin2018deep} & Lin et al.  &  & x &  & x &  & &  x & & x & 2018 \\  \hline
		\cite{216799} & eSGD  &  & x &  & x &  &  &   &  & x &  2018 \\  \hline
		\cite{DBLP:journals/corr/abs-1803-03383} & HALP  & x &  &  & x & x & x & x  & x & x &  2018 \\  \hline
		\cite{TicTac} & TicTac  &  &  & x & x &  & &  x &  &  &  2019 \\  \hline

\hline 
		
	\end{tabular} 
	\caption{Categorization of approaches on efficient communication in data-parallel training. }
	\label{tab:communicationApproaches}
	\vspace{-15pt}
\end{table}

\subsubsection{Communication}
\label{sec:data-parallel communication}

Synchronizing the model replicas in data-parallel training requires communication between workers and between workers and parameter servers (in the centralized architecture). The main challenge in optimizing the communication is to prevent that communication becomes the bottleneck of the overall training process, which would leave compute resources under-utilized.  We identified three main approaches for communication efficiency: (1) Reducing the model precision, (2) compressing the model updates, and (3) improving the communication scheduling. The current landscape of communication approaches is categorized in Table~\ref{tab:communicationApproaches}. 

\emph{(1) Reducing the model precision.} Reducing the precision of the parameters of the model saves communication bandwidth when parameter updates need to be transferred over the network. Additionally, it reduces the model size, which can be useful when the model is deployed on resource-constrained hardware such as GPUs. Precision reduction can be achieved by reducing the precision of the parameters' data types, e.g., from double precision to single floating point precision or even less.

Gupta et al.~\cite{Gupta:2015:DLL:3045118.3045303} limited the numerical precision of DL models to 16-bit fixed-point arithmetic. They found that when applying stochastic rounding as opposed to the common round-to-nearest method, the scheme with limited precision achieves nearly the same model accuracy as when applying the traditional 32-bit floating point arithmetic that is typically used in DL. This allows for reducing the model size by half. When applied to a data-parallel DL system, this will also reduce the network bandwidth needed for communicating parameter updates between workers and parameter servers; the approach itself does not depend on a specific synchronization method or parallel architecture.
DoReFa-Net~\cite{DBLP:journals/corr/ZhouNZWWZ16} by Zhou et al. focuses on CNNs. Their main idea is to reduce the numerical precision of weights, activations and gradients to different bit-widths. They report to use 1-bit weights, 2-bit acitvations and 6-bit gradients on the AlexNet CNN~\cite{Krizhevsky:2012:ICD:2999134.2999257} and still reach an accuracy that is competitive to a 32-bit representation. High-accuracy low-precision (HALP) by De Sa et al.~\cite{DBLP:journals/corr/abs-1803-03383} is an algorithm that combines two optimization techniques in order to reach high model accuracy despite of limited parameter precision. First, they use stochastic variance-reduced gradient (SVRG)~\cite{Johnson:2013:ASG:2999611.2999647} to reduce noise from gradient variance. Second, to reduce noise from parameter quantization, they introduce a new technique called \emph{bit centering}, i.e., re-centering and re-scaling of the fixed-point representation of the parameters as the model converges. Same as Gupta et al.~\cite{Gupta:2015:DLL:3045118.3045303}, they rely on stochastic rounding for quantization.

Model quantization is commonly applied to reduce the size of already trained models for more efficient inference, e.g., on mobile devises. Such \emph{post-training quantization} is implemented, e.g., in TensorFlow Lite~\cite{tf-post-training-quantization-doc}, MXNet~\cite{mxnet-model-quantify} and PyTorch~\cite{qnnpack}. Model quantization at training time is less common; it is not widely implemented in DL frameworks.

\emph{(2) Compressing the model updates.} The model updates communicated between workers and between workers and parameter servers can be compressed. Lossless compression is limited in the achievable compression rate, as redundancy in the parameter updates is typically limited. Instead, lossy compression is applied. The main methods in the literature are gradient quantization (reducing the number of bits per gradient) and gradient sparsification (communicating only important gradients that have a significant value). 

Seide et al.~\cite{seide-compress-gradients} report on quantizing the gradients in a speech DNN to \emph{one single bit}. To still achieve high accuracy, they propose a technique called \emph{error-feedback}. In error-feedback, when quantizing gradients, they save the induced quantization error and add it into the respective next batch gradient before its quantization. Hence, the gradients' information is not lost by quantization, but all gradients are eventually added into the model. TernGrad~\cite{wen2017terngrad} by Wen et al. introduces \emph{ternary gradients}, i.e., the gradient can have the value -1, 0, or 1. To improve on the model accuracy, they propose layer-wise ternarizing (i.e., using a different quantization for each layer) and gradient clipping (i.e., limit the magnitude of each gradient before quantizing it).  QSGD~\cite{qsgd-alistarh} by Alistarh et al. follows a similar approach. They apply stochastic rounding (cf. Gupta et al.~\cite{Gupta:2015:DLL:3045118.3045303} and De Sal et al.~\cite{DBLP:journals/corr/abs-1803-03383}) and statistical encoding; the key idea of the latter is that not all values are equally likely which is exploited in the encoding scheme. 

Besides quantization, another common technique is gradient sparsification. It is based on the observation that in the training process, many gradients are very small (i.e., have a value close to 0) and do not contribute much to the training. By leaving out gradients with insignificant values, the communication volume can be reduced. The parameter server by Li et al.~\cite{186214} allows for gradient sparsification via user-defined filters. 
eSGD~\cite{216799} is a gradient sparsification approach for federated architectures. Lin et al.~\cite{lin2018deep} propose a gradient sparsification approach that is based on a threshold. Only gradients larger than the threshold are transmitted. The rest of the gradients are accumulated until the threshold is reached. This is similar to the error-feedback that Seide et al.~\cite{seide-compress-gradients} proposed for quantization. Lin et al. combine their sparsification approach with \emph{momentum correction} to mitigate issues introduced by the transmission of accumulated small gradients. Further, they apply gradient clipping.

Gradient quantization and sparsification at training time are implemented in a number of open source DL frameworks. CNTK implements the 1-bit stochastic gradient descent by Seide et al.~\cite{seide-compress-gradients}. MXNet supports 2-bit quantization with error-feedback; 1-bit quantization and sparsification techniques are on the roadmap~\cite{mxnet-quantization-doc}. 

\emph{(3) Communication scheduling.} Communication patterns in data-parallel DL are typically \emph{bursty}, especially in strictly synchronous systems: All workers may share their updated parameters at the same time with their peer workers or parameter servers. To prevent that the network bandwidth is exceeded and communication is delayed, the communication of the different workers can be scheduled such that it does not overlap. Furthermore, when bandwidth is constrained, but too many parameter updates are to be sent, communication scheduling can prioritize specific messages over others, e.g., depending on freshness or on significance for the model convergence.

B\"{o}sen~\cite{Wei:2015:MCC:2806777.2806778} by Wei et al. maximizes network communication efficiency by prioritizing updates that are most significant to the model convergence. TicTac~\cite{TicTac} by Hashemi et al. is a system for communication scheduling in synchronous centralized architectures. They observe that in many ML and DL systems such as TensorFlow and PyTorch, parameters are transmitted randomly in the training and inference process. This results in high variance in iteration time, which slows down the process. To overcome that problem, TicTac enforces a schedule of network transfers that optimizes the iteration time.
MLNet~\cite{190634} by Mai et al. is a communication layer for centralized data-parallel ML. They combine a tree-shaped communication overlay with traffic control and prioritization to mitigate network bottlenecks.

Sophisticated communication scheduling algorithms have not found their way into open source DL frameworks yet. This may be due to the novelty of the methods.

\subsection{Scheduling and Elasticity}
\label{sec:scheduling}

The \emph{scheduling problem} in DL refers to how to map the (possibly parallel) DL training processes to the processing nodes in the distributed infrastructure. We identified three different aspects of scheduling in DL. 

First, there is \emph{single-tenant} scheduling (Section~\ref{sec:single-tenant}): How to map the processes (e.g., workers and parameter servers) of a single tenant, i.e., training job, to the available infrastructure? In case that mapping is dynamic, and we can change the number of training processes (e.g., number of workers and number of parameter servers) as well as the infrastructure (e.g., number of compute nodes), we also talk about \emph{elasticity} in the scheduling problem. 

Second, there is \emph{multi-tenant} scheduling (Section~\ref{sec:multi-tenant}): Given multiple competing training jobs (each having a number of processes), how to map them to the available infrastructure? The multi-tenant case introduces additional challenges such as a larger complexity and additional requirements or constraints such as fairness among the tenants. 

Third, there is a specific scheduling problem that concerns the \emph{creation} of training jobs in DL, namely, the \emph{model architecture and hyper-parameter search} (Section~\ref{sec:hyperparameter-search}). This problem is tightly coupled to single-tenant and multi-tenant scheduling.

\subsubsection{Single-tenant}
\label{sec:single-tenant}

In single-tenant scheduling, we assume a dedicated, but possibly dynamic, set of resources (compute nodes, CPUs, GPUs) that is available to host a set of processes that originate from a single DL training job. With training job, we refer to all processes involved in performing the training of a single DL model. Depending on the parallelization method, this may comprise workers that train complete (data parallelism) or partial (model parallelism) model replicas as well as parameter servers. Now, scheduling needs to answer the following questions: (1) Which process is placed on which resource (such as compute node, CPU, or GPU)? (2) When or in what order are the processes that are placed on the same resource executed? (3) When and how are the number of processes and/or resources adapted?

In model parallelism, one of the major problems to be solved is to partition the model into multiple parts. We have discussed this issue and state-of-the-art approaches for addressing it in Section~\ref{sec:parallelization methods}. Once the model is partitioned, the next important questions are \emph{where} to place the model parts and \emph{when} to train \emph{which} partition of the model. As a training iteration of a model partition can only be executed when all input data of that partition is available, there are dependencies in scheduling the different model partitions. Mayer et al.~\cite{Mayer:2017:TPS:3154842.3154843} have formalized the scheduling problem in model-parallel DL. While they propose a couple of heuristic algorithms, none of them have been implemented in the context of DL systems. In particular, there are interdependencies between the model partition and the scheduling problem, which are yet to be fully explored. Additional challenges arise with the advent of dynamic control flow~\cite{Yu:2018:DCF:3190508.3190551, Jeong:2018:IED:3190508.3190530} that renders static scheduling infeasible. Park et al.~\cite{park2019accelerated} propose \emph{layer placement}, which is however limited to CNNs. STRADS~\cite{Kim:2016:SDF:2901318.2901331} by Kim et al. is a model-parallel ML framework with an advanced scheduler. In particular, STRADS can take into account dependency structures in model partitions and is capable of prioritizing computations. To do so, the user has to implement his training task via three functions \texttt{schedule}, \texttt{update} and \texttt{aggregate}. While the paper contains example implementations of classical ML algorithms such as LASSO and topic modeling, it is not straight-forward to implement a model-parallel DL training job via the STRADS interface.

Litz~\cite{216041} by Qiao et al. is an elastic ML framework that exposes an event-driven programming model. In Litz, computations are decomposed into micro-tasks that are dynamically scheduled on a cluster. The scheduler takes into account dependencies and consistency requirements of the ML model. To enable interruption-free elasticity, the input data is ``over-partitioned'' across \emph{logical executors} which are dynamically mapped to physical resources. This allows even for transparent scaling of stateful workers, i.e., workers that keep local state that is not shared via the parameter servers or directly with peer workers. This property is useful when different model state is affected by the training of different ranges of input data, such that for faster access that portion of the model state is directly kept at the worker.

Proteus~\cite{Harlap:2017:PAM:3064176.3064182} by Harlap et al. exploits transient resources such as Amazon EC2 spot instances and Google Compute Engine preemptible instances. Its main concepts are a parameter server framework that is optimized for bulk addition and revocation of transient resources, and a resource allocation component that dynamically allocates transient resources to minimize the overall monetary cost per work based on highly dynamic spot markets.

CROSSBOW~\cite{DBLP:journals/corr/abs-1901-02244} by Koliousis et al. is a decentralized data-parallel DL system that can automatically tune the number of workers at run-time. To do so, the number of workers is increased during the training until no more increase in training throughput can be observed. This way, the available infrastructure can be utilized in an optimal way. Further, CROSSBOW comes with a dynamic task scheduler to execute workers on GPUs based on resource availability. FlexPS~\cite{Huang:2018:FFP:3187009.3177734} by Huang et al. takes on the problem of \emph{varying workloads} during the execution of ML training. As sources of varying workloads, Huang et al. mention adaptive hyper-parameters (specifically, the batch size), and advanced SGD methods such as SVRG~\cite{Johnson:2013:ASG:2999611.2999647}. As a result of this problem, the parallelism degree, i.e., the number of workers, needs to be adapted to re-balance the trade-off between communication and computation in data-parallel training.

\subsubsection{Multi-tenant}
\label{sec:multi-tenant}

In a multi-tenant environment, multiple training jobs (tenants) share a common set of resources. Hence, a \emph{resource scheduler} is responsible to schedule the processes of the different tenants on the resources. There is a large variety of general purpose resource schedulers such as Mesos~\cite{Hindman:2011:MPF:1972457.1972488}, YARN~\cite{Vavilapalli:2013:AHY:2523616.2523633}, and Borg~\cite{Verma:2015:LCM:2741948.2741964}. However, these are not tailored to the specific properties of DL training tasks. For instance, in DL, the convergence rate of a training task varies over time. Typically, in the beginning of training, progress is made very quickly; however, as training evolves over many epochs, the improvements on model accuracy decrease. Further, different DL training jobs may have very different training curves~\cite{Zhang:2017:SQS:3127479.3127490}. Taking into account these DL-specific properties allows for formulating new, DL-specific optimization goals, e.g., maximizing the overall training progress over all scheduled training jobs. Hence, new DL resource schedulers are being proposed.

Dolphin~\cite{lee2016dolphin} by Lee et al. is an elastic centralized data-parallel ML framework. In Dolphin, the configuration of the parameter servers and workers is adapted dynamically according to a cost model and continuous monitoring. Here, the configuration refers to the number of servers and workers, the distribution of training data across workers and the distribution of model parameters across parameter servers. The system is implemented on top of Apache REEF~\cite{Weimer:2015:RRE:2723372.2742793}, a framework for distributed applications.
Optimus \cite{Peng:2018:OED:3190508.3190517} by Peng et al. is a system that dynamically adjusts the number and placement of workers and parameter servers of a training job at run-time to achieve the best resource efficiency and training speed. To do so, it builds performance models based on sampling that estimate the number of training epochs needed until convergence and the impact of different configurations (number of workers and parameter servers) on the training speed. Then, a greedy algorithm computes the best allocation of resources to workers and parameter servers. Considering multiple concurrent training jobs to be scheduled, Optimus aims to minimize the average job completion time. An additional challenge tackled by Optimus is to divide the model parameters onto the parameter servers such that the load is balanced. Compared to the general-purpose scheduling policies Dominant Resource Fairness~\cite{Ghodsi:2011:DRF:1972457.1972490} and Tetris~\cite{Grandl:2014:MPC:2619239.2626334}, Optimus shows significant improvements in average job completion time and makespan\footnote{The makespan of a set of training jobs is the total time elapsed from the arrival of the first job to the completion of all jobs.}. Jeon et al.~\cite{jeon2019analysis} analyze log traces from a large-scale DL cluster system. In particular, they analyze the trade-off between locality constraints and queuing delays for large training jobs that occupy a lot of (GPU) resources. Further, they observe that co-locating different jobs on the same server may significantly impact their performance. Finally, they also analyze failures in DL training and the root causes why they occur. They differentiate between failures caused by the infrastructure, by the DL framework, and by the user. Based on their analysis, they propose a couple of best practices for multi-tenant DL scheduling. First, they emphasize that locality is a major design goal of schedulers that should definitely be taken into account. Second, they highlight that isolation of jobs is important in order to avoid performance interference. Third, they propose that new jobs should first be tested on a small dedicated set of servers before being admitted to the cluster. Ease.ml~\cite{Li:2018:ETM:3187009.3177737} is an ML service platform that employs a multi-tenant resource scheduler. Users define their training jobs in a declarative language and submit them to ease.ml via a web interface. Then, ease.ml not only schedules that job on the available resource, but also automates model architecture and hyper-parameter search. The overall goal of ease.ml is to maximize the average model accuracy achieved among all tenants, i.e., users of the system. SLAQ~\cite{Zhang:2017:SQS:3127479.3127490} by Zhang et al. has a similar goal, but supports a broader set of optimization goals. It does not only maximize average accuracy, but also solves a min-max problem to provide fairness among the tenants. Ray~\cite{Nishihara:2017:RML:3102980.3102998, 222605} from UC Berkeley is a distributed system that is specialized to support the requirements of reinforcement learning. The design of Ray makes it necessary to dynamically schedule millions of tasks per second, where each task represents a remote function invocation that may only take as little as a few milliseconds to complete. The scheduler in Ray is hierarchical with two levels: one single global scheduler and a local scheduler per node. As long as a node is not overloaded, the local scheduler schedules its tasks autonomously. However, if a local scheduler detects overload, it forwards tasks to the global scheduler, which assigns them to other nodes.

Besides publications that describe concrete multi-tenant schedulers, there are publications that describe \emph{DL services}. IBM Fabric for Deep Learning~\cite{ffdl:17} (FfDL) is a cloud-based deep learning stack used at IBM by AI researchers. Based on FfDL, IBM offers \emph{DL as a Service} (DLaaS)~\cite{DLaaS}, a fully automated cloud solution for DL. Hauswald et al.~\cite{7284053} describe Djinn, an open infrastructure for DL as a service in large-scale distributed infrastructures, as well as Tonic, a suite of DL applications for image, speech and language processing. They analyze the workloads of their system and propose a design for large-scale infrastructures that is suitable to DL workloads. One of their findings is that employing GPUs for DL training and inference can reduce total cost of ownership tremendously compared to applying only CPUs. In their analysis, they take into account upfront capital expenditures, operating costs and financing costs. While GPUs have a higher purchase price, such investment pays off due to lower operating costs when processing DL workloads.

\subsubsection{Model Architecture and Hyper-Parameter Search}
\label{sec:hyperparameter-search}

Model architecture and hyper-parameter search is a crucial problem in DL training. Given a specific task (e.g., image classification), what is the best model architecture (e.g., CNN with how many layers and what layer dimensions) that can reach the best accuracy? And what are the best hyper-parameter settings to reach model convergence quickly? Finding the answer to those questions is difficult. The typical approach is to repeatedly try out different architectures and hyper-parameter settings in order to find the best one, i.e., a search based on experimental evaluations~\cite{Sparks:2015:AMS:2806777.2806945}. The search can be random~\cite{Bergstra:2012:RSH:2503308.2188395} or guided by more sophisticated models, such as random forests and Bayesian optimization~\cite{Hutter:2014:EAA:3044805.3044891} or even reinforcement learning~\cite{baker2017designing, zoph2017neural}. What all of those methods have in common is that they repeatedly spawn new training jobs with new configurations (architectures and hyper-parameter settings) that need to be scheduled on a shared set of distributed resources. Here, we discuss scheduling approaches that explicitly take into account workloads that are generated by such search strategies.

TuPAQ~\cite{Sparks:2015:AMS:2806777.2806945} by Sparks et al. is a system for automatically generating and executing model search configurations. Based on performance profiles provided by a domain expert, TuPAQ automatically optimizes the amount of resources for data parallel training. Batching together training jobs that access the same training data reduces network load and allows for further optimizations in the execution. 
HyperDrive~\cite{Rasley:2017:HEH:3135974.3135994} by Rasley et al. is a scheduler that optimizes the hyper-parameter search more aggressively than TuPAQ does. In particular, HyperDrive supports \emph{early stopping} of the training of poorly configured jobs. Further, by incorporating the trajectory of learning curves of the trained models, HyperDrive predicts the expected accuracy improvement. Based on that, more resources are assigned to training jobs that have a high expected accuracy improvement compared to other configurations. HiveMind~\cite{accelerating-deep-learning-workloads-through-efficient-multi-model-execution} by Narayanan et al. is a system designed to optimize the execution of multiple DL training jobs on a single GPU. The system executes a batch of models jointly and performs cross-model optimizations such as operator fusion (e.g., shared layers on different model architectures) and shared I/O (e.g, using the same training data for different configurations). Gandiva~\cite{222611} by Xiao et al. is a system that schedules sets of jobs for hyper-parameter search simultaneously on a cluster of GPU-powered compute nodes. By exploiting early feedback, subsets of the jobs can be killed and resources can be freed. Based on profiling of job execution times, Gandiva employs a fine-grained application-aware time-slicing of the GPU resources to exploit them optimally. To place the jobs on GPUs, Gandiva also takes into account their memory footprint as well as communication intensity to minimize interference between job executions.

\subsection{Data Management}
\label{sec:mod}

One of the great challenges of large-scale DL is handling the data that is involved. On the one hand side, this refers to the management of training data, whose volume easily exceeds the capabilities of a single disk or multiple disks on a single server. On the other hand side, it refers to the management of the DL models, both fully trained as well as snapshots of models currently in the training phase. The training and model data need to be handled in a suitable manner, while taking into account the available distributed infrastructure, the running training processes and the resource scheduling in the data center.

\subsubsection{Training Data}
Obtaining large labeled training data sets is a hard problem. One approach to achieve this is to resort to \emph{manual labeling}. For instance, to build the ImageNet data set, the authors relied on crowd sourcing via Amazon Mechanical Turk, which led to high accuracy of the labels~\cite{5206848}. However, manual labeling is expensive and time-consuming. Hence, there are several approaches to allow for training with highly noisy training data that can be easily obtained, e.g., from web image search. Xiao et al.~\cite{Xiao_2015_CVPR} embed a label noise model into a DL framework. They train two CNNs: one of the CNNs predicts the label while the other CNNs predicts the noise type of the training data set. For training, they first pre-train both CNNs with clean training data. Then, they train the models with the noisy data, but mix in data with clean labels to prevent model drift. Overall, learning from noisy data is a vast research area (cf., e.g.,~\cite{mnih2012learning, DBLP:journals/corr/SukhbaatarF14}) which we will not cover in its entirety in this survey. 

Besides obtaining data (noisy or clean), preprocessing of the training data is an important step in data management. This includes \emph{normalization} such as cropping, resizing and other adjustments on image data~\cite{cirecsan2012multi}, or data \emph{augmentation} such as creating spectrograms from speech data~\cite{graves2014towards}. Beyond normalization and augmentation, training a DL model with \emph{distorted} training data can increase the model's robustness to noisy input data~\cite{Zheng_2016_CVPR}. Hence, preprocessing of training data takes an important role in the overall DL architecture. For instance, Project Adam and Facebook both describe that preprocessing is performed on distinct data servers~\cite{186212, 8327042}. 

Once the training data is obtained and preprocessed, it has to be provided to the training servers for feeding it into the DL models in the training iterations.  Ozeri et al.~\cite{Ozeri:2018:OSD:3286490.3286562} use simple and cheap \emph{object storage} to store and provide the training data. The shortcoming of object storage is that the bandwidth of data provisioning is limited to about 35 MB per second for a single request, while the throughput of training data on a machine with 4 GPUs can reach up to 570 GB per second according to the authors' own measurements. They add a FUSE-based file system to the DL stack which translates POSIX API requests into REST API requests.  To overcome the read throughput limitation, their storage layer converts a single read request into multiple concurrent requests to the object storage to yield higher aggregate bandwidth. Kubernetes Volume Controller~\cite{kvc} (KVC) is an advanced interface for training data management on Kubernetes clusters. It provides an abstraction on training data that can be used by the training processes, and internally manages data placement and replication transparently to the user. Hoard~\cite{DBLP:journals/corr/abs-1812-00669} by Pinto et al. is a distributed caching system that stripes the training data across local disks of the worker machines for fast access. Training data is loaded from the backend only once and can then be provisioned from the cache for subsequent epochs and across training tasks that use the same training data (e.g., at exploratory architecture and hyper-parameter search). 

\subsubsection{Model Data}

Managing the trained models is as important as the training process itself. According to Vartak et al.~\cite{Vartak:2016:MOD:2939502.2939516}, model management involves tracking, storing and indexing of trained models. The goal of model management is to facilitate the sharing, querying and analyzing of the DL models. To make that possible, there are a number of current initiatives and approaches. 

To facilitate interoperability between different DL frameworks, the Open Neural Network Exchange Format (ONNX)~\cite{onnx} is being developed. ONNX is the de-facto standard for exchange of model data between DL frameworks. DL frameworks that natively support ONNX are Caffe2, Chainer~\cite{chainer_learningsys2015, chainermn_mlsys2017}, CNTK~\cite{Seide:2016:CMO:2939672.2945397}, MXNet~\cite{mxnet_learningsys2016}, PyTorch~\cite{paszke2017automatic}, PaddlePaddle~\cite{paddlepaddle}, Matlab and SAS~\cite{sas}. Moreover, model converters are available for TensorFlow~\cite{199317}, Keras, Apple CoreML~\cite{coreML}, SciKit-learn~\cite{pedregosa2011scikit}, XGBoost~\cite{xgboost}, LIBSVM~\cite{CC01a}, and Tencent ncnn~\cite{ncnn}. 
ModelDB~\cite{Vartak:2016:MOD:2939502.2939516} by Vartak et al. is a system for model management that provides automatic tracking of ML models, indexing, and querying via SQL or via a visual interface. Beyond the models themselves, ModelDB also manages meta data (e.g., hyper-parameters of the training process), quality metrics and training and test data sets for each model. ModelHub~\cite{7930008} by Miao et al. is a system that serves a similar purpose as ModelDB. Beyond providing a versioned model storage and query engine and a domain specific language for model architecture and hyper-parameter search, ModelHub also provides a repository-based model sharing system for easy exchange of DL models between different organizations.

\section{Comparison of Deep Learning Frameworks}
\label{sec:comparison}

Since the rise of DL, a large number different DL frameworks and tools have been developed and many of them are open source. They implement different concepts of parallelization and distribution that we have discussed in Section~\ref{sec:challenges and techniques}. Having a large choice of open-source DL frameworks is one of the drivers of innovative DL research. In this section, we review and compare current open-source DL frameworks and tools. 

\subsection{Evaluation Criteria}

We discuss and compare the frameworks according to the following criteria.

\emph{(1) APIs.} DL frameworks should support a large range of programming languages, so that experts from different domains have easy access to them. Moreover, they should provide high-level abstractions so that a running DL use case can be created quickly without many obstacles.

\emph{(2) Support for distribution and parallelization.} In a cloud environment, resources are available abundantly and on demand. DL frameworks should allow for easy and intuitive support for distribution and parallelization without need for custom code. We specifically examine this point with regard to the parallelization methods and optimizations we have discussed in Section~\ref{sec:challenges and techniques}. Here, we also discuss the possibility for users to fine-tune their deployment according to their needs. This relates to the DL frameworks' support for custom definitions of the DL model and loss functions and developing custom code for parameter servers or custom topologies in decentralized systems.

\emph{(3) Community.} As the field of DL is dynamically evolving, with new DL model architectures and parallelization methods being proposed, it is crucial for a DL framework to have an active community that discusses and implements the most promising approaches. We measure community activity by the number of commits on the official Github repositories in the past six months (i.e., between October 2018 and March 2019) as well as the total number of topics with the respective tags on StackOverflow\footnote{Due to limitations of the StackOverflow search, we did not confine the search to recent topics, but report the overall numbers without time constraint.} (https://stackoverflow.com/). 

We emphasize that we do not discuss and compare the performance of DL frameworks; a comprehensive performance evaluation of DL frameworks is out of the scope of this survey article. There are other studies that compare performance, e.g., by Liu et al.~\cite{liu2018usability} or Jäger et al.~\cite{Jager:2018:PTD:3286490.3286561}.

\subsection{Detailed Analysis}

In the following, we discuss the frameworks in more detail. Table~\ref{tab:comparison} provides an overview. 

{\footnotesize
\begin{longtable}[t]{|p{1.5cm}|p{1.0cm}||p{1.8cm}|p{5.5cm}|p{1.6cm}|}
		\hline 
	Name & Papers  & API & Distribution and Parallelization  & Community \\
		\hline \hline
		Caffe & \cite{DBLP:journals/corr/JiaSDKLGGD14} & CLI, Python, Matlab & No native support for distribution. & Github: 2 \newline StOv: 2,750 \\  \hline
		Caffe2 & n/a & C++, Python & \textbullet\ Decentralized only \newline \textbullet\ Synchronous only \newline  \textbullet\ Model quantization supported \newline  \textbullet\ Gradient quantization not supported  \newline  \textbullet\ Communication scheduling not supported & Github: n/a \newline StOv: 116 \\  \hline
		Chainer &  \cite{chainer_learningsys2015, chainermn_mlsys2017} & Python & \textbullet\ Decentralized only \newline \textbullet\ Synchronous only \newline  \textbullet\ Model quantization not supported \newline  \textbullet\ Gradient quantization not supported  \newline  \textbullet\ Communication scheduling not supported  & Github: 3,939 \newline StOv: 132 \\  \hline
		CNTK & \cite{Seide:2016:CMO:2939672.2945397} & C++, C\#, Python, Brain-Script & \textbullet\ Centralized and decentralized  \newline  \textbullet\ Bounded asynchronous training via BMUF~\cite{7472805} \newline  \textbullet\ Model quantization not supported \newline  \textbullet\ 1-bit gradient quantization~\cite{seide-compress-gradients} supported \newline  \textbullet\  Communication scheduling not supported  & Github: 138 \newline StOv: 488 \\  \hline
		DL4j &  n/a & Java & \textbullet\ Centralized and decentralized  \newline  \textbullet\ Synchronous and asynchronous \newline  \textbullet\ Model quantization not supported \newline  \textbullet\ Modified 1-bit gradient quantization by Strom~\cite{strom2015scalable, dl4j-quantify} supported \newline  \textbullet\  Communication scheduling not supported & Github: 390 \newline StOv: 243 \\  \hline
		Keras &  n/a & CNTK, DL4j, TensorFlow, Theano & \textbullet\ Model quantization supported \newline  \textbullet\  Higher-level concepts must be implemented in the DL framework that employs Keras & Github: 310 \newline StOv: 14,630 \\  \hline
		MXNet &  \cite{mxnet_learningsys2016} &C++, Go, Java-Script, Julia, Matlab, Perl, Python,  R, Scala, Wolfram & \textbullet\ Centralized only \newline \textbullet\ Synchronous and asynchronous \newline  \textbullet\ Model quantization supported \newline  \textbullet\ 2-bit gradient quantization with error-feedback supported~\cite{mxnet-quantization-doc}  \newline  \textbullet\ Communication scheduling not supported  & Github: 837 \newline StOv: 455 \\  \hline
		PyTorch &  \cite{paszke2017automatic} & C++, Python &  \textbullet\ Centralized and decentralized  \newline  \textbullet\ Synchronous and asynchronous \newline  \textbullet\ Model quantization not supported \newline  \textbullet\ Gradient quantization not supported \newline  \textbullet\  Communication scheduling not supported & Github: 3,484 \newline StOv: 2,413 \\  \hline 
		SINGA & \cite{Ooi:2015:SDD:2733373.2807410} & C++, Python & \textbullet\ Centralized and decentralized \newline  \textbullet\ Synchronous and asynchronous \newline  \textbullet\ Model quantization not supported \newline  \textbullet\ Gradient quantization not supported \newline  \textbullet\  Communication scheduling not supported & Github: 44 \newline StOv: 0 \\  \hline
		TensorFlow &  \cite{199317} & C++, Go, Java, Java-Script, Python, Swift &  \textbullet\ Centralized \newline  \textbullet\ Synchronous and asynchronous \newline  \textbullet\ Model quantization supported \newline  \textbullet\ Gradient quantization not supported \newline  \textbullet\  Communication scheduling not supported  &  Github: 10,930  \newline StOv: 39,334 \\  \hline
		Theano &  \cite{bergstra2011theano} & Python & No native support for distribution. & Github: 55 \newline StOv: 2,389 \\  \hline  
		
		\multicolumn{3}{l}{} \\[10pt]
		
	\caption{Comparison of open source DL frameworks and libraries. StOv: StackOverflow.}
	\label{tab:comparison}
\end{longtable}
}

\textbf{Caffe} is a DL framework developed by Berkeley AI research and community contributors.  It comes with command line, Python and Matlab APIs. A specialty of Caffe is the \emph{model zoo}, a collection of pre-trained models for an easy start. It runs on CUDA platforms (using the cuDNN library) for easy parallelization on GPUs. Caffe does not support distributed training out-of-the-box. However, there are forks and extensions of Caffe such as Intel Caffe\footnote{https://github.com/intel/caffe} and CaffeOnSpark\footnote{https://github.com/yahoo/CaffeOnSpark} that support distributed training. There is only little information available in the Caffe documentation of how to customize the framework, e.g., to develop new loss functions. As Caffe does not support multi-node deployment, custom parallelization techniques can not be implemented either. Commit activity on Github has almost completely ceased. On StackOverflow, there are 2,750 questions tagged with ``Caffe'', a high value compared to other frameworks.

\textbf{Caffe2} is a successor of the Caffe framework developed by Facebook and community contributors. The API is available in C++ and Python. The models from Caffe can be easily converted to work with Caffe2. Beyond that, Caffe2 provides its own model zoo as well. Caffe2 extends Caffe in the following points. First of all, Caffe2 naturally supports distributed training. There is native support for decentralized data-parallel training using the synchronous model; there is no support for (bounded) asynchronous training and no parameter server architecture. There is also native support for quantized models, i.e., models with reduced data type precision. Recently, the code of Caffe2 has been merged into PyTorch. This makes it hard to assess the update frequency of the Caffe2 code. On StackOverflow, there are 116 questions tagged with ``Caffe2'', a rather low value compared to other frameworks. 

\textbf{Chainer} is  a DL framework developed by the Japanese company Preferred Networks with several industrial partners and community contributors. It is written in Python and only has a Python interface. There is good documentation on how to write custom functions, optimizer, and trainers. \emph{ChainerMN} is an extension package that enables distributed and parallel DL on multiple nodes. It supports data parallelism via a decentralized all-reduce architecture using the synchronous training method (no parameter server or asynchronous training are supported). There were 3,939 commits to the official Github repository in the past six months, which is a comparably high value. On StackOverflow, there are 132 questions tagged with ``Chainer'', a rather low value compared to other frameworks.

\textbf{CNTK} (Microsoft Cognitive Toolkit) is a DL framework developed by Microsoft and community contributors. The API is available in C++, C\# and Python. Additionally, CNTK provides a custom model description language called BrainScript. The model evaluation function can also be used from Java programs. Data-parallel and distributed training is supported out-of-the-box. The 1-bit stochastic gradient descent by Seide et al.~\cite{seide-compress-gradients} is integrated into the framework. CNTK supports the centralized architecture with parameter servers, using asynchronous training or blockwise model update and filtering (BMUF)~\cite{7472805}, a variant of bounded asynchronous training.  Currently, model parallelism is not supported by CNTK. Extending CNTK is easy. New operators, loss functions, etc. can be implemented with an API. There were 138 commits to the official Github repository in the past six months, which is a comparably low value. On StackOverflow, there are 488 questions tagged with ``CNTK'', an average value compared to other frameworks.

\textbf{Deeplearning4j} is a DL framework developed by the company Skymind and community contributors organized in the Eclipse foundation. The framework is written in Java and C++ (for core components), and the API is available in Java which makes it accessible for Java, Scala and Clojure projects (but not from Python). It supports distributed and parallel training by using Spark. There are two variants of data-parallel training implemented. First, a decentralized asynchronous approach proposed by Strom~\cite{strom2015scalable} that also incorporates quantization of gradients. Second, centralized synchronous training with a single parameter server. There is no support for model parallelism. It is easily possible to create custom layer implementations, but more sophisticated customization (loss functions, parallelization configurations, etc.) is not supported. There were 390 commits to the official Github repository in the past six months, which is an average value. On StackOverflow, there are 243 questions tagged with ``Deeplearning4j'', a rather low value compared to other frameworks.

\textbf{Keras} is not a DL framework, but a DL library that can be integrated into many other DL frameworks, such as CNTK, Deeplearning4j, TensorFlow and Theano. It is developed as a community project, initiated by F. Chollet. Keras is written in Python which allows for its easy integration into other Python-based frameworks. Parallel training on GPUs is naturally supported; higher-level parallelization concepts must be implemented to the DL framework that uses Keras. Model quantization (to 8-bit model weights) is supported directly in Keras. The library is easily extensible with new modules. There were 310 commits to the official Github repository in the past six months, which is an average value. On StackOverflow, there are 14,630 questions tagged with ``Keras'', a very high value compared to other frameworks.

\textbf{MXNet} is a DL framework and an Apache project (incubating). Its API is available for C++, Python, Julia, Matlab, JavaScript, Go, R, Scala, Perl, and Wolfram Language. MXNet supports a wide range of parallelization approaches. Model parallelism is supported for multiple GPUs on a single node; there is no support for multi-node model parallelism though. Data parallelism is realized via the centralized architecture with support for using multiple parameter servers via a sharded key-value store. Both synchronous and asynchronous training are supported out-of-the-box. MXNet also supports post-training 8-bit model quantization tailored to the Intel(R) Math Kernel Library for Deep Neural Networks (Intel(R) MKL-DNN)~\cite{mxnet-model-quantify}. In the training process, 2-bit gradient quantization with error-feedback is supported~\cite{mxnet-quantization-doc}. It is easy to implement custom operators or layers as well as loss functions. There were 837 commits to the official Github repository in the past six months, which is an average value.  On StackOverflow, there are 455 questions tagged with ``MXNet'', an average value compared to other frameworks.

\textbf{PyTorch} is a DL framework developed by Facebook and community contributors. Its API is available for C++ and Python. PyTorch has native support for distributed, data-parallel training, as well as model-parallel training. For data-parallel training, PyTorch implements the decentralized architecture and supports synchronous as well as asynchronous training. PyTorch supports model quantization via the QNNPACK library~\cite{qnnpack}. Gradient quantization is not supported out-of-the-box. Writing new operators or layers is easily done via extending an interface; it is also possible to write custom loss functions. There were 3,484 commits to the official Github repository in the past six months, which is a comparably high value.  On StackOverflow, there are 2,413 questions tagged with ``PyTorch'', a rather high value compared to other frameworks.

\textbf{SINGA} is a DL framework and Apache project (incubating) which is developed by community contributors. The initiators of the project are from the National University of Singapore. It has APIs in C++ and Python. Singa has native support for distributed, data-parallel and model-parallel training, as well as hybrid parallelism (combining data and model parallelism). Data parallelism is implemented via the centralized approach with support for multiple parameter servers. However, the decentralized architecture can be emulated by employing each worker with a local parameter server. Both synchronous and asynchronous training are supported. There is no support for model or gradient quantization. Customization is more difficult than in the other frameworks: The documentation does not contain any hints on how to implement custom layers or loss functions. There were 44 commits to the official Github repository in the past six months, which is a comparably low value. On StackOverflow, there are no questions tagged with ``Singa'' or ``Apache Singa'', and only one single question is returned when searching for the keyword ``Singa''.

\textbf{TensorFlow} is an ML framework developed by Google and community contributors. The API is available for C++, Go, Java, JavaScript, Python and Swift. Additionally, the community offers bindings for C\#, Haskell, Ruby, Rust and Scala. TensorFlow natively supports distributed and parallel training. In particular, it supports both model parallelism and data parallelism. In data parallelism, the centralized approach via parameter servers is supported, using either asynchronous or synchronous training. Trained models can be quantized using TensorFlow Lite~\cite{tf-post-training-quantization-doc}. Currently, there is no native support for gradient quantization or communication scheduling. Customization of layers and loss functions is straight forward via implementing the available interfaces. There were 10,930 commits to the official Github repository in the past six months, which is an extremely high value. On StackOverflow, there are 39,334 questions tagged with ``TensorFlow'', which is the highest number among all analyzed DL frameworks.

\textbf{Theano} is a DL framework developed by Montreal Institute for Learning Algorithms at the Universit\'{e} de Montr\'{e}al. The API is available only for Python. There is no support for distributed training on multiple nodes. However, using multiple GPUs on a single node is supported. Theano supports model parallelism, but no data parallelism. New layers can be implemented via an interface. It is also possible to define custom loss functions. At the time of writing this survey, commits to the official Github repository have a low frequency. According to a posting on the Theano mailing list\footnote{https://groups.google.com/forum/\#!topic/theano-users/7Poq8BZutbY}, major development of Theano ceased with the release of version 1.0.; however, new maintenance releases have been issues since then. There were still 55 commits to the official Github repository in the past six months. On StackOverflow, there are 2,389 questions tagged with ``Theano'', a rather high value compared to other frameworks.

\textbf{Others.} There are a couple of other frameworks that we do not cover in detail in our comparison for various reasons.  Minerva~\cite{wang2014minerva} is an open sourced DL system, but has not been maintained for the past 4 years. SparkNet~\cite{moritz2015sparknet} allows for distributed DL on Spark, but has not been maintained for the past 3 years. Neon~\cite{neon} is another DL framework that has ceased development for more than 1 year. Scikit-learn~\cite{pedregosa2011scikit} is an ML framework and it is not specific to DL. While neural network training is implemented, there is no support for using GPUs or distributed training. The Weka workbench~\cite{Frank2010} is a collection of ML and data mining algorithms. WekaDeeplearning4j~\cite{wekadl4j} is a DL package for the Weka workbench. As backend, it uses Deeplearning4j, which we have discussed above.


\section{Conclusions and Outlook}
\label{sec:conclusion}

DL is becoming increasingly important in industry and academia and is without doubt one of the most impactful revolutions in computer science in the past years. However, the rapid pace in which the field is developing makes it difficult to keep an overview. In particular, DL is currently investigated from many different perspectives and in different communities. In this survey, we took a deeper look into DL from the perspective of \emph{scalable distributed systems}. We investigated the main challenges to make DL systems scale, and have reviewed the common techniques that have been proposed by researchers to tackle those challenges. This included an analysis of the distributed infrastructures used in DL training as well as techniques for parallelization, scheduling and data management. Finally, we provided an overview and comparison of the current open-sourced DL systems and tools, and analyzed which of the techniques developed in research have actually been implemented. We saw that the wide range of techniques for scalable DL are implemented in open-source DL frameworks. This shows that there is a fruitful interaction between research and practical applications which is one of the reasons why DL has gained such a large momentum. 

We can draw from our survey a couple of insights on how to design future DL infrastructures and tools. In our opinion, management of training and model data becomes a larger challenge with the proliferation of more training data and more DL models. This demands better tool support such that new bottlenecks and limitations for DL scalability can be mitigated. Furthermore, current developments and advances in decentralized training, e.g., federated learning, may change the requirements and design of DL infrastructures and tools. If the infrastructure becomes more heterogeneous, this must be reflected in DL tools that can not only just deal with such heterogeneity, but even exploit it to optimize the training process. 

Looking into the future, we see a couple of trends that will be important in the next years. While research on scalable DL was mostly focused on the parallelization and distribution aspects of DL training, there is a need to investigate other parts of the DL environment, such as data management and multi-tenant scheduling. This is a large field for research in the distributed systems and database community. Furthermore, DL serving, i.e., using trained DL models for inference, receives growing attention~\cite{201468, Gujarati:2017:SDA:3135974.3135993, 8360337}. Although DL serving is closely related to DL training, the requirements and, hence, the solutions are totally different. Another important aspect of DL is privacy~\cite{Shokri:2015:PDL:2810103.2813687, Abadi:2016:DLD:2976749.2978318, LI201776}, which receives growing attention due to an increasing awareness in the society for privacy issues in the era of Big Data, fueled by legislative reforms such as the General Data Protection Regulation (GDPR) in the European Union. There is an interesting trade-off between the ever-increasing demand for more training data to improve DL models and the principle of data avoidance and data economy to protect privacy.

\bibliography{references}
\bibliographystyle{acm}

\end{document}